\documentclass[11pt]{article}
\usepackage{axodraw}
\usepackage{epsfig}
\usepackage{amsfonts}
\usepackage{bbm}
\newcommand{\la}[1]{\label{#1}}
\newcommand{\be}{\begin{equation}}
\newcommand{\ee}{\end{equation}}
\newcommand{\ba}{\begin{eqnarray}}
\newcommand{\ea}{\end{eqnarray}}
\newcommand{\rmi}[1]{{\mbox{\scriptsize #1}}}
\newcommand{\rmii}[1]{{\mbox{\tiny\rm{#1}}}}
\newcommand{\fig}{Fig.~}
\newcommand{\eq}{Eq.~}
\newcommand{\se}{Sec.~}
\newcommand{\eqs}{Eqs.~}
\newcommand{\nr}[1]{(\ref{#1})}

\newcommand{\nn}{\nonumber \\}
\newcommand{\fr}[2]{{\frac{#1}{#2}\,}}
\newcommand{\msbar}{{\overline{\mbox{\rm MS}}}}

\renewcommand{\vec}[1]{{\bf #1}}
\newcommand{\tinymsbar}{{\overline{\mbox{\tiny\rm{MS}}}}}
\newcommand{\Lambdamsbar}{{\Lambda_\tinymsbar}}
\newcommand{\Nf}{N_{\rm f}}

\newcommand{\rmO}{{\mathcal{O}}}
\newcommand{\bmu}{\bar\Lambda} %{\bar\mu}

\newcommand{\bM}{\beta_\rmi{M}}

\newcommand{\aE}[1]{\alpha_\rmi{E#1}}
\newcommand{\aEms}[1]{\alpha_\rmi{E#1}^\tinymsbar}
\newcommand{\bE}[1]{\beta_\rmi{E#1}}
\newcommand{\bEms}[1]{\beta_\rmi{E#1}^\tinymsbar}
\newcommand{\logT}{\ln\frac{\bmu}{4 \pi T}}
\newcommand{\logz}[1]{\frac{\zeta'(-#1)}{\zeta(-#1)}}
\newcommand{\gammaE}{\gamma_\rmii{E}}
\def\lsi{\raise0.3ex\hbox{$<$\kern-0.75em\raise-1.1ex\hbox{$\sim$}}}
\def\gsi{\raise0.3ex\hbox{$>$\kern-0.75em\raise-1.1ex\hbox{$\sim$}}}

\newcommand{\Tint}[1]{{\hbox{$\sum$}\!\!\!\!\!\!\!\int\,}_{\!\!\!\!\raise-0.9ex\hbox{$\scriptstyle{#1}$}}}

\newcommand{\ZZ}{{\mathbb{Z}}}

\newcommand{\e}{\epsilon}
\renewcommand{\(}{\left(}
\renewcommand{\)}{\right)}
\renewcommand{\[}{\left[}
\renewcommand{\]}{\right]}
\def\sumint{\hbox{$\sum$}\!\!\!\!\!\!\!\int}

\newcommand{\pic}[1]{\;\parbox[c]{30pt}{\begin{picture}(30,30)(0,0)
\SetWidth{1.0}\SetScale{1.0} #1 \end{picture}}\;}

\newcommand{\picb}[1]{\;\parbox[c]{45pt}{\begin{picture}(45,30)(0,0)
\SetWidth{1.0}\SetScale{1.0} #1 \end{picture}}\;}

\def\Asc(#1,#2)(#3,#4,#5){\CArc(#1,#2)(#3,#4,#5)}
\def\Lsc(#1,#2)(#3,#4){\Line(#1,#2)(#3,#4)}
\def\scfc{0.7}  % picture scale factor 
\def\phgt{21}   % all picture height 30 * \scfc
\def\pwc{21}    % c   picture width  30 * \scfc
\def\pwcb{31.5} % cb  picture width  45 * \scfc
\newcommand{\PIC}[4]{\;\parbox[c]{#2 pt}{\begin{picture}(#2,#3)(0,0)
\SetWidth{2.0}\SetScale{#4} #1 \end{picture}}\;}
\renewcommand{\pic}[1]{\PIC{#1}{\pwc}{\phgt}{\scfc}}
\renewcommand{\picb}[1]{\PIC{#1}{\pwcb}{\phgt}{\scfc}}

\def\TopoSB(#1,#2,#3){\picb{#1(0,15)(7.5,15) #2(22.5,15)(15,0,180)%
 #3(22.5,15)(15,180,360) #1(37.5,15)(45,15)}}
\def\TopoST(#1,#2){\picb{#1(0,0)(22.5,0) #1(22.5,0)(45,0)%
 #2(22.5,15)(15,-90,270)}} 
\def\ToptSS(#1,#2,#3,#4){\picb{#1(0,15)(7.5,15) #1(37.5,15)(45,15)%
 #4(7.5,15)(37.5,15) #2(22.5,15)(15,0,180) #3(22.5,15)(15,180,360)}}
\def\ToprSTT(#1,#2,#3){\picb{#1(0,0)(22.5,0) #1(22.5,0)(45,0)%
 #2(22.5,15)(15,-90,90) #2(22.5,15)(15,90,270)% 
 #3(22.5,35)(5,-90,270)}}
\def\TopoSX(#1,#2,#3){\picb{#1(0,30)(7.5,15) #1(0,0)(7.5,15)%
 #2(22.5,15)(15,0,180)  #3(22.5,15)(15,180,360)% 
#1(37.5,15)(45,30) #1(37.5,15)(45,0)}}

\makeatletter \@addtoreset{equation}{section} \makeatother
\renewcommand{\theequation}{\arabic{section}.\arabic{equation}}
\makeatletter
\renewcommand\section{\@startsection {section}{1}{\z@}%
                                   {-5.5ex \@plus -1ex \@minus -.2ex}% bfr-skip
                                   {2.3ex \@plus.2ex}%
                                   {\normalfont\large\bfseries}}
\renewcommand\subsection{\@startsection{subsection}{2}{\z@}%
                                     {-3.25ex\@plus -1ex \@minus -.2ex}%
                                     {1.5ex \@plus .2ex}%
                                     {\normalfont\normalsize\bfseries}}
\renewcommand\thesection {\@arabic\c@section}
\renewcommand\thesubsection   {\thesection.\@arabic\c@subsection}
\renewcommand{\@seccntformat}[1]{%
\csname the#1\endcsname.\hspace{1.0em}}
\makeatother
\begin{document}

\begin{titlepage}
\begin{flushright}
BI-TP 2007/05\\
ECT*-07-06\\
hep-ph/0703307\\
\end{flushright}
\begin{centering}
\vfill

{\Large{\bf Four-loop pressure of massless O($N$) scalar field theory}}

\vspace{0.8cm}

A.~Gynther$^\rmi{a}$, %%\footnote{\tt gynthera@brandonu.ca},
M.~Laine$^\rmi{b}$, %%\footnote{\tt laine@physik.uni-bielefeld.de},
Y.~Schr\"oder$^\rmi{b}$,  %%\footnote{\tt yorks@physik.uni-bielefeld.de},
C.~Torrero$^\rmi{b}$, %%\footnote{\tt torrero@physik.uni-bielefeld.de},
A.~Vuorinen$^\rmi{c}$ %%\footnote{\tt vuorinen@phys.washington.edu}

\vspace{0.8cm}

${}^{\rm a}${\em
Dept of Physics \& Astronomy, Brandon University, 
Brandon, Manitoba, R7A 6A9 Canada\\}

\vspace*{0.2cm}

${}^{\rm b}${\em
Faculty of Physics, University of Bielefeld, 
D-33501 Bielefeld, Germany\\}

\vspace*{0.2cm}

${}^{\rm c}${\em
Dept of Physics, University of Washington, 
Seattle, WA 98195--1560, USA\\}

\vspace*{0.8cm}
 
\mbox{\bf Abstract}

\end{centering}

\vspace*{0.3cm}
 
\noindent
Inspired by the corresponding problem in QCD, we determine
the pressure of massless O($N$) scalar field theory up to order
$g^6$ in the weak-coupling expansion, where $g^2$ denotes the quartic 
coupling constant. This necessitates the computation of all 4-loop vacuum 
graphs at a finite temperature: by making use of methods developed 
by Arnold and Zhai at 3-loop level, we demonstrate that this task is 
manageable at least if one restricts to computing the logarithmic terms 
analytically, while handling the ``constant'' 4-loop contributions 
numerically.  We also inspect the numerical convergence of the weak-coupling 
expansion after the inclusion of the new terms. Finally, we point out that 
while the present computation introduces strategies that should be 
helpful for the full 4-loop computation on the QCD-side, 
it also highlights the need to develop novel computational 
techniques, in order to be able to complete this formidable task
in a systematic fashion. 
\vfill
\noindent
 
%\noindent
%PACS numbers: 
%11.10.Wx, %        Finite temperature field theory
%11.15.Bt, %        General properties of perturbation theory
%11.15.Ha, %        Lattice gauge theory
%12.38.Bx, %        Perturbative calculations in QCD
%98.80.Cq  %        Early Universe
%\\
%Keywords:

\vspace*{1cm}
 
\noindent
March 2007

\vfill

\end{titlepage}

%%%%%%%%%%%%%%%%%%%%%%%%%%% SECTION %%%%%%%%%%%%%%%%%%%%%%%%%%%%%%%%%%%%%
%
\section{Introduction}

Motivated for instance by hydrodynamic studies
of heavy ion collision experiments, and dark 
matter relic density computations in cosmology, 
a lot of theoretical work has been devoted to the perturbative determination 
of the pressure of hot QCD in recent years. 
As a result of 3-loop and 4-loop computations, 
corrections to the non-interacting Stefan-Boltzmann
law have been determined up to relative orders 
% $\rmO(g^2)$~\cite{es}, 
% $\rmO(g^3)$~\cite{jk}, 
% $\rmO(g^4\ln(1/g))$~\cite{tt}, 
$\rmO(g^4)$~\cite{az}, 
$\rmO(g^5)$~\cite{zk,bn2}, and 
$\rmO(g^6\ln(1/g))$~\cite{gsixg}, 
where $g$ denotes the renormalized strong coupling constant.
The first presently unknown order, $\rmO(g^6)$, 
contains a non-perturbative 
coefficient~\cite{linde,gpy}, but that can also 
be estimated numerically~\cite{plaq,nspt_mass}.
All orders of $g$ are available in the formal limit of 
large $\Nf$~\cite{Nf}, where $\Nf$ counts 
the number of massless quark flavours.
Similar results have also been obtained for the case of non-zero
quark chemical potentials at finite~\cite{av2} and large~\cite{Nfmu}
$\Nf$. Moreover, first
steps towards the inclusion of finite quark masses, important
for phenomenological applications, have been taken~\cite{massdep}. 
Finally, coefficients up to the order $\rmO(g^5)$ are available
even for the standard electroweak theory, at temperatures 
higher than the electroweak scale~\cite{gv}. 

Conceptually, it would be quite desirable to extend
these results up to the full order $\rmO(g^6)$. 
The reason is that this is the first order where at least 
the leading contributions from the various momentum
scales relevant for hot QCD, $k\sim 2\pi T, gT, g^2 T$, 
have been fully accounted for.  As mentioned, the non-perturbative
input needed to describe the effect of the softest momenta
$k \sim g^2 T$ is already available~\cite{plaq,nspt_mass}. Moreover, 
several perturbative contributions of $\rmO(g^6)$ are known: 
in the notation of Ref.~\cite{gsixg}, 
$\bM$~\cite{aminusb} (which accounts for the scales $k\sim gT$), 
$\bE{2}$~\cite{bE2}, 
$\bE{3}$~\cite{gE2}, 
as well as $\bE{4}$ and $\bE{5}$~\cite{quart} have been computed. 
Nevertheless,  a single coefficient, $\bE{1}$, 
coming from the hard scales $k\sim 2\pi T$, remains undetermined.  

The recipe to compute $\bE{1}$ is in principle simple: 
it is defined to be the ``naive'' (i.e.\ unresummed) 4-loop
contribution to the pressure, computed by regulating all 
divergences via dimensional regularization in $d=4-2\epsilon$ dimensions. 
This computation contains  infra\-red (IR)
divergences, which manifest themselves
as an uncancelled $1/\epsilon$-divergence 
in the final (renormalized) result. 
This IR pole gets only cancelled once the contributions of the soft
modes are properly resummed, a step that has already been completed, and
produces another $1/\epsilon$-pole~\cite{gsixg}, with the opposite sign. 

Unfortunately, the practical implementation of this 
computation is far from trivial. In fact, 
to show that it is feasible at all, it is the purpose
of the present paper to demonstrate that the 4-loop order
can at least be reached in models somewhat simpler than QCD. 

The model that we consider is scalar field theory, 
with a global O($N$) symmetry, i.e.\ the ``$\lambda \phi^4$''-theory. 
To keep the analogy with QCD in mind, we follow here the frequent 
convention of denoting $\lambda \equiv g^2$. The pressure
of this theory has been computed to high orders in $g$
in parallel with that for QCD: the orders 
$\rmO(g^4)$~\cite{fst,az}, 
$\rmO(g^5)$~\cite{ps,bn1}, 
and
$\rmO(g^6\ln(1/g))$~\cite{bn1}
have been reached. 
The order $\rmO(g^6)$ involves a coefficient analogous to $\bE{1}$, 
which again contains an uncancelled $1/\epsilon$-pole. 
It is the goal of the present paper to compute this $\bE{1}$ (as well as
all other relevant coefficients), 
showing that the pole cancels, and thus to determine the full 
pressure up to $\rmO(g^6)$, at any finite $N$.

It is perhaps worth stressing
that even though the technical challenges addressed in this paper
have a direct counterpart in QCD, there is of course the conceptual difference
that in our model the order $\rmO(g^6)$ contains no non-perturbative
coefficients, since the scale $k\sim g^2 T$ does not exist in 
scalar field theory. 
Therefore no lattice studies
of the type in Refs.~\cite{plaq,nspt_mass} need to be invoked here. 

The plan of this paper is the following. 
We start by carrying out the naive 4-loop computation
of the pressure  in \se\ref{se:betaE1}, leading to the 
determination of the coefficient $\bE{1}$. 
In~\se\ref{se:setup}, we elaborate on how the naive computation
can be re-interpreted and 
incorporated in a proper setting such that all  
divergences cancel. This leads to our final finite result. We 
discuss various formal and numerical aspects of the result
in \se\ref{se:results}, and conclude in~\se\ref{se:concl}.

%%%%%%%%%%%%%%%%%%%%%%%%%%% SECTION %%%%%%%%%%%%%%%%%%%%%%%%%%%%%%%%%%%%%
%
\section{Naive 4-loop computation}
\la{se:betaE1}

Our starting point is the bare theory
\be
 \mathcal{L}_\rmi{B} = 
 \fr12 \sum_{\mu = 1}^{4-2\epsilon} \sum_{i=1}^N
 \partial_\mu\phi_i \partial_\mu\phi_i 
 + 
 \frac{1}{4!} g^2_\rmi{B} \Lambda^{2\epsilon} \Bigl(
 \sum_{i=1}^N \phi_i \phi_i 
 \Bigr)^2
 \;, \la{LB}
\ee
where $\Lambda$ is the scale parameter introduced in connection
with dimensional regularization. 
We work in Euclidean metric throughout.
The fields $\phi_i$ have the 
dimension $[\mbox{GeV}]^{1-\epsilon}$, while the bare coupling
$g_\rmi{B}^2$ is dimensionless. 
The theory can be renormalized in the $\msbar$ scheme by setting
\be
 g^2_\rmi{B} = g^2 +  
 \frac{g^4 }{(4\pi)^2} \frac{\beta_1}{\epsilon} +   
 \frac{g^6 }{(4\pi)^4}
 \biggl(
  \frac{\beta_1^2}{\epsilon^2} + \frac{\beta_2}{2\epsilon} 
 \biggr)
 + \rmO(g^8)
 \;, \la{gB2}
\ee
where (see, e.g.,\ Ref.~\cite{bmk})
\be
 \beta_1 = \frac{N+8}{6}
 \;, \quad
 \beta_2 = -\frac{3 N + 14}{6}
 \;. \la{betas}
\ee

The observable that we are interested in, is the pressure, or minus 
the free energy density, of this theory, as function of the temperature $T$
and the renormalized coupling constant $g$. Formally, the pressure
is given by the path integral
\be
 p(T) \equiv \lim_{V\to\infty} \frac{T}{V} 
 \ln \int \! \mathcal{D} \phi_i \, \exp(- S_\rmi{B})
 \;, \la{fullpdef}
\ee
where $V$ is the volume, $V = \int \! {\rm d}^{3-2\epsilon} \vec{x}$, 
and $S_\rmi{B}$ is the bare action, 
$ 
 S_\rmi{B} = \int_0^{\beta} \! {\rm d}\tau 
 \int \! {\rm d}^{3-2\epsilon} \vec{x} \, \mathcal{L}_\rmi{B}
 \;.
$
The temperature $T$ is given by $T=1/\beta$, and the path integral
is taken with the usual periodic boundary conditions over 
the $\tau$-direction. Note that in the presence of dimensional
regularization, the pressure thus defined has the dimension 
$[\mbox{GeV}]^{4-2\epsilon}$.

%%%%%%%%%%%%%%%%%%%%%%%%%%%%%%%%%%%%%%%%%%%%%%%%%%%%%%%%%%%%%%%%%%%%%%%%%%
%%%%%%%%%%%%%%%%%%%%%%%%%%%%%%%%%%%%%%%%%%%%%%%%%%%%%%%%%%%%%%%%%%%%%%%%%%

Going into momentum space, we define the usual (sum-)integrals as
\ba
 \int_p &\equiv& \Lambda^{2\e}\int\! 
 \fr{{\rm d}^{3-2\e} \vec{p}}{(2\pi)^{3-2\e}} 
 \;\;=\;\; \(\!\fr{e^{\gammaE}\bar{\Lambda}^2}{4\pi}\!\!\)^{\!\!\e}\!\!\int\!
 \fr{{\rm d}^{3-2\e} \vec{p}}{(2\pi)^{3-2\e}},
 \label{asdfg}\\
 \sumint_{P} &\equiv& T \sum_{p_0} \int_p
 \;,
\ea
where $p_0$ stands for bosonic Matsubara momenta, 
$p_0 = 2 \pi n T$, $n \in \ZZ$. 
Furthermore, $\Lambda$ is the MS scale parameter,  
while $\bmu$ is the $\msbar$ one; these are related through
$\bmu^2 \equiv 4 \pi \Lambda^2 \exp(-\gammaE)$.
We then define
\ba
 {\cal I}^m_{n} &\equiv& \sumint_P \fr{(p_0)^m}{\(P^2\)^n}
 \;, \la{Imn} \\
 \Pi(P)&\equiv&\sumint_Q \fr{1}{Q^2(Q-P)^2}
 \;, \la{Pi} \\
 \bar{\Pi}(P)&\equiv&\sumint_Q \fr{1}{Q^4(Q-P)^2}
 \;. \la{bPi}
\ea

%%%%%%%%%%%%%%%%%%%%%%%%%%%%%% FIGURE %%%%%%%%%%%%%%%%%%%%%%%%%%%%%
\begin{figure}[t]
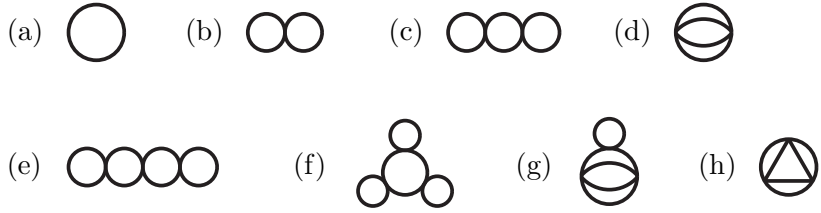


% \centerline{\epsfxsize=11cm\epsfysize=3.2cm \epsfbox{phi4pic1.ps}}

\centering \ba \nonumber
\begin{array}{llll}
({\rm a})&~\!\!\!\!\pic{\CArc(15,15)(15,0,360)} 
 \quad\;\;\;({\rm b})\;\; \pic{\CArc(10,15)(10,0,360)\CArc(30,15)(10,0,360)}
 \quad\;\;\;\;\;\;({\rm c})\;\; \pic{\CArc(10,15)(10,0,360)% 
 \CArc(30,15)(10,0,360)%
 \CArc(50,15)(10,0,360)}\;\;\;\;
 \;\;\;\;\;\;\;\;\,({\rm d})\;\;
 \pic{\CArc(15,15)(15,0,360)\CArc(15,2)(20.2,41,139)%
 \CArc(15,28)(20.2,221,319)}
 \nn \nn\nn ({\rm e})&~\!\!\!\!\pic{\CArc(10,15)(10,0,360)%
 \CArc(30,15)(10,0,360)\CArc(50,15)(10,0,360)\CArc(70,15)(10,0,360)}
 \;\;\;\;\;\;\;\;\;\;\;\;\;\;\;\;\;\;\;\;({\rm f})\;\;\;\;\;\;
 \pic{\CArc(12,12)(12,0,360)\CArc(12,32)(8,0,360)%
 \CArc(-5.33,2)(8,0,360)\CArc(29.33,2)(8,0,360)}
 \quad\;\;\;\;\;({\rm g})\;\;\, \pic{\CArc(15,10)(15,0,360)%
 \CArc(15,-3)(20.2,41,139)\CArc(15,23)(20.2,221,319)%
 \CArc(15,33)(8,0,360)}
 \quad\;\;\;({\rm h})\;\; \pic{\CArc(15,15)(15,0,360)%
 \Line(2,7.5)(28,7.5)\Line(2,7.5)(15,30)\Line(15,30)(28,7.5)}
\end{array}
\ea

\caption[a]{\it The 1-, 2-, 3- and 4-loop full theory diagrams 
contributing to the pressure.
    \label{fig:graphs}
    }
\end{figure}
%%%%%%%%%%%%%%%%%%%%%%%%%%%%%%%%%%%%%%%%%%%%%%%%%%%%%%%%%%%%%%%%%%%%

Up to 4-loop order, the strict loop expansion for the pressure 
of our theory contains the graphs  (a)--(h) in \fig\ref{fig:graphs}, 
which we denote by the symbols $I_\rmi{a}$--$I_\rmi{h}$. Denoting 
furthermore ${\cal I}_n \equiv {\cal I}^0_n$, and employing the 
sum-integrals defined in \eqs\nr{Imn}--\nr{bPi}, we can immediately
write down their expressions in the forms:
\ba
I_\rmi{a} &=& N\fr{\pi^2}{90}T^4 \, [ 1 + \rmO(\epsilon) ]
\;, \\
I_\rmi{b} &=& -\fr{N(N+2)}{24}g_\rmi{B}^2({\cal I}_{1})^2
\;,\\
I_\rmi{c} &=& \fr{N(N+2)^2}{144}g_\rmi{B}^4({\cal I}_{1})^2{\cal I}_2
\;,\\
I_\rmi{d} &=& \fr{N(N+2)}{144}g_\rmi{B}^4\sumint_P \[\Pi(P)\]^2
\;,\\
I_\rmi{e} &=& -\fr{N(N+2)^3}{864}g_\rmi{B}^6({\cal I}_{1})^2({\cal I}_2)^2
\;,\\
I_\rmi{f} &=& -\fr{N(N+2)^3}{1296}g_\rmi{B}^6({\cal I}_{1})^3{\cal I}_3
\;,\\
I_\rmi{g} &=& -\fr{N(N+2)^2}{216}g_\rmi{B}^6
               {\cal I}_{1}\sumint_P \Pi(P)\,\bar{\Pi}(P)
\;,\\
I_\rmi{h} &=& -\fr{N(N+2)(N+8)}{1296}g_\rmi{B}^6\sumint_P \[\Pi(P)\]^3
\;.
\ea
Taking now into account the renormalization of the coupling constant 
up to 2-loop order, we must replace the bare coupling $g_\rmi{B}^2$
by the renormalized one, $g^2$, through \eq\nr{gB2}. 
Recalling that a weak-coupling expansion
(i.e.,\ an expansion in $g^2$) of the path-integral in \eq\nr{fullpdef}
does not coincide with the loop expansion at finite temperatures, 
because of well-known IR divergences in multiloop graphs, 
we will denote the result for the sum of these graphs in the 
following by $p_\rmi{E}(T)$, in contrast to the physical 
pressure denoted by $p(T)$; for the latter we assume that a consistent 
evaluation has been carried out to a certain order in $g^2$, 
irrespective of how many loop orders this takes.
Summing together the graphs shown leads then to the  
following expression for  $p_\rmi{E}(T)$:
\ba
 \Lambda^{2\e} p_\rmi{E} 
 &=& 
 N\fr{\pi^2}{90}T^4 \, [ 1 + \rmO(\epsilon) ]
 -g^2\times\fr{N(N+2)}{24}({\cal I}_{1})^2
 + \nn &+&
 g^4\times\fr{N(N+2)}{144}
 \Bigg\{({\cal I}_{1})^2\bigg[(N+2){\cal I}_2-\fr{N+8}{(4\pi)^2\e}\bigg]
 +\sumint_P \[ \Pi(P) \]^2\Bigg\}
 - \nn &-&
 g^6\times\fr{N(N+2)}{1296}\Bigg\{(N+2)^2\({\cal I}_1\)^3{\cal I}_3
% + \nn & &  
 + \fr{3}{2}({\cal I}_{1})^2\bigg[(N+2)^2\({\cal I}_2\)^2
 - \nn & & 
 - \fr{2(N+2)(N+8)}{(4\pi)^2\e}{\cal I}_2
 +\fr{1}{(4\pi)^4}\biggl(\fr{(N+8)^2}{\e^2} -\fr{3(3N+14)}{\e}\biggr)\bigg]
 + \nn & &
 + 6(N+2){\cal I}_1\sumint_P \Pi(P)\bar{\Pi}(P)
 +(N+8)\sumint_P \[ \Pi(P) \]^3
 -\fr{3(N+8)}{(4\pi)^2\e}\sumint_P \[ \Pi(P) \]^2\Bigg\}
 \;. \nn \la{fullp}
\ea

From Ref.~\cite{az}, we can immediately read off the results 
for most of the sum-integrals above, utilizing the formulae
\ba
 && \hspace*{-1.0cm}
 {\cal I}^m_n =2^{m-2n+1}\pi^{m-2n+3/2}T^{m-2n+4}
 \(\fr{\Lambda^2}{\pi T^2}\)^{\!\!\e}
 \fr{\Gamma (n-3/2+\e)}{\Gamma (n)}\zeta (2n-m-3+2\e)
 \;, \label{in}\\
 && \hspace*{-1.0cm}
 \sumint_P \[ \Pi(P)\]^2
 =\fr{1}{(4\pi)^2}\(\fr{T^2}{12}\)^{\!\!2}\bigg\{\fr{6}{\e}
 +36\,\ln\,\fr{\bar{\Lambda}}{4\pi T}+48\fr{\zeta '(-1)}{\zeta(-1)}
 -12\fr{\zeta '(-3)}{\zeta(-3)}
 % \nn &&\hspace{4.6cm}
 +\fr{182}{5}+\rmO(\e)\bigg\}
 \;. \nn \la{az}
\ea
This leaves us with the task of evaluating the 
new 3- and 4-loop sum-integrals
\ba
 S_1 & \equiv & \sumint_P \Pi(P)\bar{\Pi}(P)
 \;, \la{S1def} \\
 S_2 & \equiv & \sumint_P 
 \bigg\{\[\Pi(P)\]^3-\fr{3}{(4\pi)^2\e}\[\Pi(P)\]^2\bigg\}
 \;, \la{S2def}
\ea
where we have grouped together two terms in $S_2$ 
for computational convenience.

%%%%%%%%%%%%%%%%%%%%%%%%%%%%%%%%%%%%%%%%%%%%%%%%%%%%%%%%%%%%%%%%%%%%%%%%%%
%%%%%%%%%%%%%%%%%%%%%%%%%%%%%%%%%%%%%%%%%%%%%%%%%%%%%%%%%%%%%%%%%%%%%%%%%%

The sum-integrals in \eqs\nr{S1def}, \nr{S2def} can be evaluated
through a very tedious if in principle straightforward 
application of the procedures and techniques 
that were pioneered by Arnold and Zhai in Ref.~\cite{az}.
A detailed explanation of the steps that we have taken can be found 
in Appendix~A. Here we simply quote the final results of that analysis:
\ba
 S_1 
 &=&\fr{T^2}{8(4\pi)^4}\Bigg\{\fr{1}{\e^2}
 +\fr{1}{\e}\bigg[3\,\ln\fr{\Lambda^2}{4\pi T^2}+\fr{17}{6}
 +\gammaE+2\fr{\zeta'(-1)}{\zeta(-1)}\bigg] 
 + \nn &+&
 \fr{9}{2}\(\ln\fr{\Lambda^2}{4\pi T^2}\)^2
 +\[\fr{17}{2}+3\gammaE+6\fr{\zeta '(-1)}{\zeta(-1)}\]
 \ln\fr{\Lambda^2}{4\pi T^2}+48.797635359976(4)  \Bigg\} + \rmO(\e)
 \;, \nn \la{S1res} 
 \\ 
%%%%%%%%
 S_2
 &=&
 -\fr{T^4}{16(4\pi)^4}\Bigg\{\fr{1}{\e^2}
 +\fr{1}{\e}\bigg[2\,\ln\fr{\Lambda^2}{4\pi T^2}+\fr{10}{3}
 -2\gammaE+4\fr{\zeta'(-1)}{\zeta(-1)}\bigg]
  + \nn &+&
 \(\ln\fr{\Lambda^2}{4\pi T^2}\)^{\!\!2}
 +\[\fr{6}{5}-2\gammaE+4\fr{\zeta '(-3)}{\zeta(-3)}\]
 \ln\fr{\Lambda^2}{4\pi T^2}
 -25.705543194(2)\Bigg\}
 - \nn &-&
 \fr{T^4}{512(4\pi)^2}\Bigg\{\fr{1}{\e}+4\,\ln\fr{\Lambda^2}{4\pi T^2}
 +28.92504950930(1)\Bigg\}
 + \rmO(\e) \;.  \la{S2res}
\ea
In $S_2$, we have on purpose separated one contribution 
(coming from the part denoted by $S_2^\rmi{I,b}$ in Appendix A, 
cf.\ \eq\nr{S2Ib}) 
from the rest, as it contains an IR singularity, 
which will not be cancelled by the renormalization 
of the coupling constant in the full theory, 
but only by the ultraviolet (UV) singularities originating from 
the ``soft'' contributions to the pressure (cf.\ next section).
The numbers in parentheses estimate the numerical 
uncertainties of the last digits shown.  

In order to present the full result, we introduce the following notation
for the contributions of various orders to $p_\rmi{E}$: 
\ba
 \Lambda^{2 \epsilon} p_\rmi{E} & \equiv & T^4 \Bigl[\aE{1}
 + g^2 
 %\Bigl(
 \aE{2} 
 % + {\cal O}(\epsilon)\Bigr) \nn 
 % & & \hspace*{0.7cm}
 + \frac{g^4}{(4\pi)^2} 
 % \Bigl(
 \aE{3} %+ {\cal O}(\epsilon)\Bigr) 
 + \frac{g^6}{(4\pi)^4} 
 %\Bigl(
 \bE{1} %+ {\cal O}(\epsilon)\Bigr) 
 + \rmO(g^8,\epsilon)
 \Bigr]
 \;. \la{pE}  
\ea
Here the coefficients $\aE{i},\bE{1}$ have been defined
in analogy with Ref.~\cite{gsixg}. They are dimensionless
functions of the temperature and of the regularization scale. 
Inserting \eqs\nr{in}, \nr{az}, \nr{S1res} and \nr{S2res} 
into \eq\nr{fullp}, and using everywhere the $\msbar$ scheme
scale parameter $\bmu$, we obtain the following expressions
for these coefficients: 
\ba
 \aE{1} \!\! & = &  \frac{N \pi^2}{90}
 \;, \la{aE1} \\
 \aE{2} \!\! & = & 
 -\frac{N(N+2)}{3456}
 \;, \la{aE2} \\
 \aE{3} \!\! & = &
 \frac{N(N+2)}{10368}
 \biggl[
  (N+8) \logT + (N+2) \gammaE + \frac{31}{5} + 12 \logz{1} -6 \logz{3}
 \biggr]
 \;, \la{aE3} \\  
 \bE{1} \!\! & = & \!\! \frac{N(N+2)(N+8)}{41472} \frac{\pi^2}{\epsilon} 
 - \nn \!\! & - & \!\!
 \frac{N(N+2)}{31104} %{3 (12)^4}
 \biggl\{
  (N+8)^2 \biggl( \logT \biggr)^{\!\! 2} 
 + % \nn & + & 
  N^2 \biggl[2 \gammaE \logT + \gammaE^2 + \frac{\zeta(3)}{36} \biggr]
 + \nn \!\! & + & \!\! 
 16 N \biggl[ 
 \biggl(
   \frac{107}{80} + \frac{5\gammaE}{4} - \frac{3\pi^2}{8} + \fr32\logz{1}
   - \fr34\logz{3} 
 \biggr) \logT - 3.9753932807(2)
 \biggr]  
 + \nn \!\! & + & \!\!  
 64 
 \biggl[ 
 \biggl(
   \frac{353}{160} + \frac{\gammaE}{2} - \frac{3\pi^2}{4} + 3\logz{1}
   - \fr32\logz{3} 
 \biggr) \logT -10.6970470860(3)
 \biggr]
 \biggr\}
 \;.  \nn \la{bE1}
\ea
The expressions for $\aE{1}, \aE{2}, \aE{3}$ agree with 
the results of Ref.~\cite{bn1} for $N=1$.

We note from the first line of \eq\nr{bE1} that the fully 
renormalized result indeed contains an uncancelled $1/\epsilon$ pole. 
Therefore, something must be wrong with the naive computation 
that we have carried out. We now turn to the correct procedure
for determining $p(T)$ up to $\rmO(g^6)$.

%%%%%%%%%%%%%%%%%%%%%%%%%%% SECTION %%%%%%%%%%%%%%%%%%%%%%%%%%%%%%%%%%%%%
%
\section{Resummed 4-loop computation}
\la{se:setup}

The reason that the computation carried out in the previous section
leads to a divergent result, is that it ignores the fact that certain
subsets of higher loop graphs amount to generating radiatively a mass 
for the fields $\phi_i$. In the presence of a mass, the results for 
IR sensitive loop integrals change. This cures the IR problem 
and leads to the correct weak-coupling expansion. A systematic way
to implement such a mass resummation (as well as other resummations, 
for instance for the quartic coupling) goes via effective field 
theory methods, as we now review. 

At high temperatures, the relevant effective field theory framework
is that of dimensional reduction~\cite{dr,generic}. The basic observation
is that the non-zero Matsubara modes are heavy, and certainly cannot 
cause any IR divergences. Thereby
the computation of all static thermodynamic observables can be 
factorised into two parts: to the contribution from the ``hard'' 
momentum modes, $k\sim 2\pi T$, and from the ``soft'' modes, $k\sim g T$
(the scale $k\sim g^2 T$ does not appear in scalar field theory).
For the pressure we 
will denote the two parts as $p_\rmi{hard}$ and $p_\rmi{soft}$, 
respectively. The effective theory determining $p_\rmi{soft}$ reads
\be
 \mathcal{L}_\rmi{E} = 
 \fr12 \sum_{j = 1}^{3-2\epsilon} \sum_{i=1}^N
 \partial_j\phi_i \partial_j\phi_i 
 +
 \fr12 m_\rmi{E}^2 \sum_{i=1}^N \phi_i \phi_i
 +
 \frac{1}{4!} g^2_\rmi{E} \Lambda^{2\epsilon} \Bigl(
 \sum_{i=1}^N \phi_i \phi_i 
 \Bigr)^2
 + \ldots \;. \la{LE}
\ee
A series of infinitely many higher dimensional operators has been truncated 
given that, as power-counting arguments show (cf.\ Ref.~\cite{gsixg}),
they cannot contribute to the pressure at $\rmO(g^6)$. 
The theory in \eq\nr{LE} describes the dynamics of the Matsubara zero modes, 
and thus lives in three dimensions. The parameters here are to 
be understood as bare parameters. In dimensional regularization, 
the dimension of $\phi_i$ is $[\mbox{GeV}]^{1/2-\epsilon}$
and that of $g_\rmi{E}^2$ is $[\mbox{GeV}]^{1}$.
Note that for simplicity 
we have used the same notation for the fields in \eqs\nr{LB}, \nr{LE}, 
even though they are independent integration variables, with 
a different mass dimension.
The use of the subscript E is meant to keep in mind the analogy with 
the effective theory called EQCD in the context of QCD~\cite{bn2}.
The determination of the effective parameters 
(or ``matching coefficients'') in \eq\nr{LE} 
up to 2-loop level dates back to Ref.~\cite{pert}, where 
the symmetry-breaking phase transition in the case of a massive
scalar field was considered; the application of this theory 
to the computation of the pressure at very high temperatures, 
where the zero-temperature 
scalar mass can be ignored, as is done in this paper, 
was first pursued systematically in Ref.~\cite{bn1}, as far as we know.

The factorization statement now reads that the physical 
pressure can be written as~\cite{bn1} 
\be
 p(T) = p_\rmi{E}(T) + p_\rmi{M}(T) = 
 p_\rmi{hard}(T) + p_\rmi{soft}(T)
 \;,  \la{psum}
\ee
where 
\be
 p_\rmi{M}(T) \equiv  
 \lim_{V\to\infty} \frac{T}{V} 
 \ln \int \! \mathcal{D} \phi_i \, \exp(- S_\rmi{E})
 \;, \la{pM}
\ee
and 
$ 
 S_\rmi{E} = 
 \int \! {\rm d}^{3-2\epsilon} \vec{x} \, \mathcal{L}_\rmi{E}
 \;.
$
We adopt a notation in the following whereby the matching 
coefficient $p_\rmi{E}$ is a bare quantity, like $m_\rmi{E}^2$
and $g^2_\rmi{E}$, while $p_\rmi{hard}$ is defined
as its $\msbar$ scheme version. Similarly, 
$p_\rmi{soft}$ is defined to be the $\msbar$ version of $p_\rmi{M}$.

Now, if we were not to carry out any resummation --- that is, 
if the mass parameter $m_\rmi{E}^2$ in \eq\nr{LE} were
ignored as was the case in the previous section --- 
then the path integral in \eq\nr{pM} would 
vanish order by order in dimensional regularization, 
because the propagators appearing in the computation 
would contain no mass scales. Therefore, according to 
\eq\nr{psum}, the computation of the full pressure without
any resummation, produces directly the function $p_\rmi{E}$. 
In other words, the proper interpretation for the result
of the previous section is to treat $p_\rmi{E}(T)$ in \eq\nr{pE} 
as an UV matching coefficient: it contains
the contributions to the physical pressure from the hard modes, 
$k\sim 2\pi T$. Interpreted this way, it is IR finite, 
because the soft contribution $p_\rmi{M}(T)$ has been 
subtracted (which happens automatically in dimensional regularization
by ignoring the mass parameter $m_\rmi{E}^2$~\cite{bn1}).  

Our task in 
the remainder of this section is to properly compute $p_\rmi{M}$, as defined 
by \eq\nr{pM}. This result can be extracted 
directly from Ref.~\cite{aminusb}.
Setting $g\to 0$, $\lambda\to g^2_\rmi{E}/6$, 
and $d_A\to N$ there, we obtain
\ba
 \frac{\Lambda^{2\epsilon} p_\rmi{M}}{T} & = & 
 \frac{m_\rmi{E}^3}{4\pi} 
  \frac{N}{3} 
  \biggl[ 
    1 + \epsilon \biggl( 
      \ln\frac{\bmu^2}{4 m_\rmi{E}^2} + \fr83
   \biggr)  + \rmO(\epsilon^2)
  \biggr]
 - \nn & - & 
 \frac{ g^2_\rmi{E} m_\rmi{E}^2 }{(4\pi)^2}
 \frac{N(N+2)}{24}
 \biggl[ 
   1 + 2 \epsilon \biggl( 
     \ln\frac{\bmu^2}{4 m_\rmi{E}^2} + 2
   \biggr) + \rmO(\epsilon^2)
 \biggr] 
 - \nn & - & 
 \frac{g^4_\rmi{E} m_\rmi{E} }{(4\pi)^3}
 \frac{N(N+2)}{144}
 \biggl[ 
   \frac{1}{\epsilon} + 
     3 \ln\frac{\bmu^2}{4 m_\rmi{E}^2} + 8 -4 \ln 2
     -\frac{N+2}{2}
     + \rmO(\epsilon)
 \biggr] 
 + \nn  & + & 
 \frac{ g^6_\rmi{E} }{(4\pi)^4}
 \frac{N(N+2)}{1728}
 \biggl\{
  (N+2) \biggl[ \frac{1}{\epsilon} + 
   4 \ln\frac{\bmu^2}{4 m_\rmi{E}^2} + 4 - 4 \ln 2 \biggr]
   -\frac{(N+2)^2}{3} 
  - \nn  & - &
 (N+8) \frac{\pi^2}{24}
 \biggl[
  \frac{1}{\epsilon} +  
 4 \ln\frac{\bmu^2}{4 m_\rmi{E}^2} + 2 + 4 \ln 2
 - 84 \frac{\zeta(3)}{\pi^2} 
 \biggr]  + \rmO(\epsilon)
 \biggr\}
 + \rmO\Bigl( \frac{g_\rmi{E}^8}{m_\rmi{E}} \Bigr)
 \;. \hspace*{0.5cm} \la{pMbare}
\ea

%%%%%%%%%%%%%%%%%%%%%%%%%%%%% FIGURE %%%%%%%%%%%%%%%%%%%%%%%%%%%%%%%%%%%%%
\begin{figure}[t]
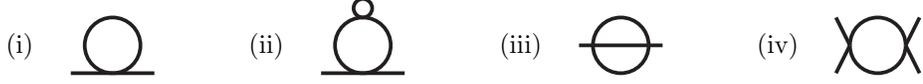


\begin{eqnarray*}
%
% 1loop
%
\mbox{\small (i)} 
\quad
\TopoST(\Lsc,\Asc) \qquad\quad
\mbox{\small (ii)} 
\quad
\ToprSTT(\Lsc,\Asc,\Asc) \qquad\quad
\mbox{\small (iii)} 
\quad
\ToptSS(\Lsc,\Asc,\Asc,\Lsc)  \qquad\quad
\mbox{\small (iv)} 
\quad
\TopoSX(\Lsc,\Asc,\Asc) % \qquad \;.
\end{eqnarray*}

\caption[a]{\it The 1-loop and 2-loop graphs needed 
for determining the matching coefficients $m_\rmi{E}^2$ and 
$g_\rmi{E}^2$ in \eqs\nr{mE}, \nr{lE}, respectively.}  
\label{fig:mass}

\end{figure}
%%%%%%%%%%%%%%%%%%%%%%%%%%%%%%%%%%%%%%%%%%%%%%%%%%%%%%%%%%%%%%%%%%%%%%%%%%

We next have to determine the values of $m_\rmi{E}^2$
and $g_\rmi{E}^2$ that appear in \eq\nr{pMbare}.
Following the notation in Ref.~\cite{gsixg}, these matching
coefficients can be written as 
\ba
 m_\rmi{E}^2 & = & T^2 \Bigl\{ g^2 
 \Bigl[ \aE{4} + 
 \aE{5} \epsilon + {\cal O}(\epsilon^2) \Bigr]
 + \frac{g^4}{(4\pi)^2} 
 \Bigl[ \aE{6} + \bE{2} \epsilon + 
 {\cal O}(\epsilon^2) \Bigr] + {\cal O}(g^6) \Bigr\}
 \;, \hspace*{0.5cm}
 \la{mE} \\
 g^2_\rmi{E} & = & T \Bigl\{ g^2 + \frac{g^4}{(4\pi)^2}  
 \Bigl[ \aE{7} +
 \bE{3} \epsilon + {\cal O}(\epsilon^2) \Bigr]  
 + {\cal O}(g^6) \Bigr\}
 \;. 
 \la{lE}  
\ea
Up to the orders indicated they are produced by the graphs
in \fig\ref{fig:mass}.
Using methods explained in some detail in Ref.~\cite{gE2},
we obtain the values
\ba
 \aE{4} & = & 
 \frac{N+2}{72} 
 \;, \la{aE4} \\ 
 \aE{5} & = & 
 \frac{N+2}{36} 
 \biggl[
   \logT 
 + 1 
 + \frac{\zeta'(-1)}{\zeta(-1)} 
 \biggr]
 \;, \la{aE5} \\ 
% \aE{6} & = & 
% \frac{N+2}{72}
% \biggl[
%  \frac{1}{\epsilon} + 
%  \frac{4-N}{3} \logT
% + 2 
% -\frac{N+2}{3} \gammaE
% + 2 \frac{\zeta'(-1)}{\zeta(-1)} 
%  \biggr]
% \;, \la{aE6} \\ 
 \aE{6} & = & 
 \frac{N+2}{72}
 \biggl[
  \frac{1}{\epsilon}
  - \frac{N}{3} \biggl( \logT + \gammaE \biggr) 
  + \frac{4}{3} \biggl( \logT
 + \fr32 
 -\frac{\gammaE}{2}
 + \fr32 \frac{\zeta'(-1)}{\zeta(-1)} \biggr) 
  \biggr]
 \;, \la{aE6} \\ 
 \aE{7} & = & 
 -\frac{N+8}{3} 
 \biggl[ \logT + \gammaE \biggr]
 \;, \la{aE7} \\ 
 \bE{2} & = & -\frac{N+2}{1728}
 \biggl\{ 
   24 N \ln^2 \frac{\bmu}{4\pi T} 
  + \nn & + & 
  16 N \biggl[ 
    \biggl( 1 + 2 \gammaE + \logz{1}  \biggr) \logT 
 + \frac{\pi^2}{16} + \gammaE + \gammaE \logz{1} - \gamma_1
   \biggr] 
 - \nn & - & 
 64 \biggl[ 
 \biggl( 1 - \gammaE + \logz{1} 
  \biggr) \logT 
  + \fr32 + \frac{\pi^2}{16} - \frac{\gammaE}{2} - \frac{\gammaE}{2} \logz{1}
  + \frac{\gamma_1}{2}
 + \nn & + &
  \fr32\logz{1} + \fr34 \frac{\zeta''(-1)}{\zeta(-1)}
 \biggr]  
%   16 \biggl[ 2(N+2) \gammaE + 
%      (N-4) \biggl( 1 + \frac{\zeta'(-1)}{\zeta(-1)} \biggr) \biggr] 
%     \logT
% + \nn & + & 
% (N-4) \pi^2 + 16 (N+2) \gammaE 
% \biggl( 1 + \frac{\zeta'(-1)}{\zeta(-1)} \biggr)
% - \nn & - & 
%  16 \biggl[
%    6 + (N+2) \gamma_1 + 6  \frac{\zeta'(-1)}{\zeta(-1)} + 
%    3 \frac{\zeta''(-1)}{\zeta(-1)}
% \biggr]
  \biggr\}
 \;, \la{bE2} \\ 
 \bE{3} & = & 
 -\frac{N+8}{3}\biggl[
 \ln^2 \frac{\bmu}{4\pi T}  + 2 \gammaE 
 \logT
 + \frac{\pi^2}{8} - 2 \gamma_1
 \biggr] 
 \;. \la{bE3} 
\ea
Here $\gamma_1$ is a Stieltjes constant, defined through
the series $\zeta(s) = 1/(s-1) + 
\sum_{n = 0}^\infty \gamma_n (-1)^n (s - 1)^n/n!$.
The values of $\aE{4}, \aE{6}$ 
agree with the results of Ref.~\cite{bn1} for $N=1$. 
The value of $\aE{7}$ agrees with the results of Ref.~\cite{pert},
for $N=1,2,4$.\footnote{%
  There is a sign error in \eq(60) 
  of Ref.~\cite{pert}, relevant for the case $N=1$.
  } 
In the following, we denote by $\aEms{i}, \bEms{i}$ couplings
from which the $1/\epsilon$-divergences have been subtracted; 
this is in fact relevant only for $\aE{6}$ and $\bE{1}$, 
cf.\ \eqs\nr{bE1}, \nr{aE6}. 

We note from \eqs\nr{mE}, \nr{aE6} that the mass parameter
$m_\rmi{E}^2$ has a divergent part. From the point of view of 
the effective theory, this divergence acts as a counterterm,
\be
 \delta m_\rmi{E}^2 = 
 \frac{g^4 T^2}{(4\pi)^2}
 \frac{N+2}{72\epsilon}
 \;. \la{ct}
\ee
Indeed, this agrees with the counterterm that can be determined 
within the effective theory, by just requiring renormalizability 
of $\mathcal{L}_\rmi{E}$~\cite{pert}. It will be convenient for
the following to also define finite parameters from which 
the divergence as well as terms 
proportional to $\epsilon$ have been subtracted; we denote these by
\ba
 \hat m_3^2(\bmu) \; \equiv \;  
 \frac{m_3^2(\bmu)}{T^2} & \equiv & 
 g^2 \, \aE{4} + \frac{g^4}{(4\pi)^2} \aEms{6}
 \;, \la{m32} \\ 
 \hat g^2_3 
 % \; \equiv \; 
 % \frac{g^2_3}{T}
  & \equiv & 
 g^2 +  \frac{g^4}{(4\pi)^2} \aE{7}
 \;. \la{lambda3}
\ea
The parameter $\hat g^2_3$ is renormalization group (RG) invariant up 
to the order computed, while $\hat m_3^2(\bmu)$ has dependence at 
order $g^4$, because the counterterm in \eq\nr{ct}
has been subtracted. 

Re-expanding now the bare expression of \eq\nr{pMbare} by treating 
the counterterm in \eq\nr{ct}
as a perturbation, we obtain a ``renormalized'' 
expression for $p_\rmi{M}$:
\be
 \frac{\Lambda^{2\epsilon} p_\rmi{M} }{T^4}
 = 
 - \frac{g^6}{(4\pi)^4}
 \frac{N(N+2)(N+8)\pi^2}{41472\epsilon} + 
 \frac{\Lambda^{2\epsilon} p_\rmi{soft} }{T^4}
 \;,  \la{pMps}
\ee
where, after sending $\epsilon\to 0$, 
\ba
 \frac{p_\rmi{soft} }{T^4} & = &
 \frac{\hat{m}_3^3(\bmu)}{4\pi} \frac{N}{3} 
 - \nn & - & 
 \frac{\hat g^2_3\hat{m}_3^2(\bmu)}{(4\pi)^2}
 \frac{N(N+2)}{24} 
 - \nn & - & 
 \frac{\hat g^4_3\hat{m}_3(\bmu)}{(4\pi)^3}
 \frac{N(N+2)}{288}
 \biggl[
   8 \ln\frac{\bmu}{m_3(\bmu)} 
  + 10 - 16 \ln 2- N 
 \biggr]   
 + \nn & + & 
 \frac{\hat g^6_3}{(4\pi)^4}
 \frac{N(N+2)}{864}
 \biggl\{
   (N+2) \biggl[ 
     2 \ln\frac{\bmu}{m_3(\bmu)} + 1 - 4 \ln 2
   \biggr]
   - \frac{(N+2)^2}{6} 
  - \nn & - & 
  (N+8) \frac{\pi^2}{24}
  \biggl[ 
   4 \ln\frac{\bmu}{m_3(\bmu)} + 1 - 2 \ln 2- 42 \frac{\zeta(3)}{\pi^2}
  \biggr]
 \biggr\}
 + \rmO\biggl(\frac{\hat g^8_3 }{ \hat{m}_3(\bmu) } \biggr)
\;. \la{psoft}
\ea 
We note from \eq\nr{pMps} 
that the only divergence appearing in $p_\rmi{M}$ after the re-expansion
of $m_\rmi{E}^2$ is of $\rmO(g^6)$. 
Therefore, none of the coefficients 
$\aE{5}, \bE{2}, \bE{3}$ in \eqs\nr{mE}, \nr{lE}, 
which multiply terms of order $\rmO(\epsilon)$, 
play a role in $p_\rmi{soft}$ at $\rmO(g^6)$. 
In this respect, the present theory differs from QCD, 
where the corresponding coefficients do play a role; 
the reason for the difference is that in QCD there is an 
$1/\epsilon$-divergence in the second term of \eq\nr{pMbare}.

To conclude this section, we remark that 
it may be convenient, following Ref.~\cite{massdep}, to also 
express $p_\rmi{E}$ in terms 
of the parameter $\hat g^2_3$ . 
This implements a certain (arbitrary) resummation; 
the practical effect is minor 
(in fact, for the scale
choice that we will make in \eq\nr{Lopt}, 
which leads to the vanishing of $\aE{7}$, 
there is no effect at all, cf.\ \eq\nr{phard}), 
and we present this last step only 
in order to allow for a more compact numerical handling of
the result. From \eqs\nr{pE}, \nr{bE1}, 
we obtain
\be
 \frac{\Lambda^{2\epsilon} p_\rmi{E}}{T^4}
 = 
 \frac{g^6}{(4\pi)^4}
 \frac{N(N+2)(N+8)\pi^2}{41472\epsilon} + 
 \frac{\Lambda^{2\epsilon} p_\rmi{hard}}{T^4}
 \;,  \la{pEph}
\ee
where, after setting $\epsilon\to 0$,  
the finite function $p_\rmi{hard}$ can be written as
\ba
 \frac{p_\rmi{hard}}{T^4} & = & \aE{1}
 + \hat g^2_3 \aE{2}
 + \frac{\hat g^4_3}{(4\pi)^2} 
 \Bigl(\aE{3} - \aE{2}\aE{7} \Bigr) 
 +  \nonumber \\[1mm]  & + & 
 \frac{\hat g^6_3}{(4\pi)^4} 
 \Bigl[
 \bEms{1} + 2 \aE{2} \aE{7}^2 - 2 \aE{3} \aE{7}
 \Bigr]
 + \rmO(\hat g^8_3)
 \;. \hspace*{0.9cm} \la{phard}
\ea

In the numerical results of the next section, 
we refer to the various orders of 
the weak-coupling expansion  according to 
the power of $\hat m_3, \hat g_3$ that appear, with 
the rule $\rmO(\hat m_3) = \rmO(\hat g_3) = \rmO(g)$. 
In other words, ``$\rmO(g^n)$'' 
denotes $\rmO(\hat g_3^{n-k} \hat m_3^{k})$ 
in the expression constituted by the sum of \nr{psoft}, \nr{phard}.
If $\hat m_3, \hat g_3$ were to be re-expanded 
in terms of $g$, one would recover the strict
weak-coupling expansion (given in \eq\nr{strict} below); 
however, 
it is useful to keep the result in an unexpanded form, because 
this makes it more manageable, and because the unexpanded form
introduces resummations of higher order contributions which 
may be numerically significant for the slowly convergent
part $p_\rmi{soft}$ in \eq\nr{psoft}~\cite{pert,bn1,bn2,gsixg,bir}
(in practice, though, the effects caused by this resummation
are not dramatic).

%%%%%%%%%%%%%%%%%%%%%%%%%%% SECTION %%%%%%%%%%%%%%%%%%%%%%%%%%%%%%%%%%%%%
%
\section{Results and discussion}
\la{se:results}

Inserting \eqs\nr{mE}, \nr{lE} into \eq\nr{pMbare}, 
summing together with \eq\nr{pE}, 
and sending $\epsilon\to 0$, we obtain the strict 
weak-coupling expansion for the pressure of our theory: 
\ba
 \frac{p(T)}{T^4} & = & 
 \aE{1} + g^2\, \aE{2} 
 + \nn  & + & 
 \frac{g^3}{4\pi} \frac{N}{3} \aE{4}^{3/2} 
 + \nn  & + & 
 \frac{g^4}{(4\pi)^2}
 \biggl[
  \aE{3} - \frac{N(N+2)}{24} \aE{4} 
 \biggr]
 + \nn  & + & 
 \frac{g^5}{(4\pi)^3}
 \frac{N}{2} \aE{4}^{1/2}
 \biggl[
  \aEms{6} - 
  \frac{N+2}{144}
  \biggl(
    8 \ln\frac{\bmu}{g T \sqrt{\aE{4}}}
   + 10 - 16 \ln 2 - N   
  \biggr)
 \biggr]
 + \nn  & + & 
 \frac{g^6}{(4\pi)^4}
 \biggl\{ 
  \bEms{1} - \frac{N(N+2)}{24}
  \biggl[ \aEms{6} + \aE{4} \aE{7}
 - \nn & - &
  \frac{N+2}{36}
  \biggl( 
    2 \ln\frac{\bmu}{g T \sqrt{\aE{4}}}
   + 1 - 4 \ln 2
  \biggr) 
  + \frac{(N+2)^2}{216}
 + \nn & + & 
  \frac{(N+8)\pi^2}{864}
 \biggl(
 % \frac{1}{\epsilon} + 
 4 \ln\frac{\bmu}{g T \sqrt{\aE{4}}}
   + 1 - 2 \ln 2 
   - 42 \frac{\zeta(3)}{\pi^2} 
 \biggr) \biggl] 
 \biggr\}
 + \rmO(g^7)
 \;. \la{strict}
\ea
We note that the $1/\e$-divergences in \eqs\nr{pMps}, \nr{pEph}
have cancelled against each other, as must be the case for a consistently
computed physical quantity. 
Inserting the expansions from \eqs\nr{aE1}--\nr{bE1}
and \eqs\nr{aE4}--\nr{aE7}, finally yields
the explicit expression
\ba
 \frac{p(T)}{T^4} & = & \frac{\pi^2 N}{90}\sum_{i=0}^6 p_i 
 \biggl( \frac{g}{4\pi} \biggr)^i
 \;, \la{complete}
\ea
where $g\equiv [g^2(\bmu)]^{1/2}$, 
and the coefficients read
\ba
 p_0 & = & 1
 \;, \la{p0} \\[2mm] 
 p_1 & = & 0 
 \;, \la{p1} \\[2mm] 
 p_2 & = &
 -\frac{5}{12} (N+2) 
 \;, \la{p2} \\ 
 p_3 & = & 
 \frac{5\sqrt{2}}{9} (N+2)^\fr32
 \;, \la{p3} \\ 
%  p_4 & = & 
%  \frac{5}{36}(N+2)
%  \biggl\{
%    (N+8) \logT 
%   + \nn &  & +   
%    N(\gammaE - 6)+ 
%    8 \biggl[ 
%    -\frac{29}{40} + \frac{\gammaE}{4} + \fr32 \logz{1} - \fr34 \logz{3}
%    \biggr] 
%  \biggr\} 
%  \;, \la{p4} \\ 
 p_4 & = & 
 \frac{5}{36}(N+2)
 \biggl\{
   N \biggl[ \logT  + \gammaE - 6 \biggr]
  + \nn &  & +   
   8 \biggl[ \logT
   -\frac{29}{40} + \frac{\gammaE}{4} + \fr32 \logz{1} - \fr34 \logz{3}
   \biggr] 
 \biggr\} 
 \;, \la{p4} \\ 
%  p_5  &= & 
%  -\frac{5}{9\sqrt{2}} (N+2)^\fr32
%  \biggl\{
%   (N+8) \logT - 12 \ln\biggl( \frac{g}{\pi} \sqrt{\frac{N+2}{72}} \biggr)
%   + \nn & & +
%     N \biggl(\gammaE - \fr32 \biggr) 
%   + 8 \biggl[ 
%    \fr98 + \frac{\gammaE}{4} -\fr34\logz{1}
%   \biggr]
%  \biggr\}
%  \;, \la{p5} \\  
 p_5  &= & 
 -\frac{5}{9\sqrt{2}} (N+2)^\fr32
 \biggl\{
   - 12 \ln\biggl( \frac{g}{\pi} \sqrt{\frac{N+2}{72}} \biggr)
  + \nn & & +
    N \biggl[\logT + \gammaE - \fr32 \biggr] 
  + 8 \biggl[ \logT + 
   \fr98 + \frac{\gammaE}{4} -\fr34\logz{1}
  \biggr]
 \biggr\}
 \;, \la{p5} \\  
%  p_6 & = & 
%  -\frac{5}{108} (N+2) 
%  \biggl\{
%   (N+8)^2 \biggl( \logT  \biggr)^2 
%   + \nn & & +
%   \biggl[ N^2(2 \gammaE - 12) 
%   + % \nn &  & + 
%   16 N \biggl(
%    -\frac{493}{80} + \frac{5\gammaE}{4} + \fr32\logz{1}-\fr34\logz{3} 
%   \biggr)
%   + \nn &  & + 
%   64 \biggl( 
%  -\frac{127}{160} + \frac{\gammaE}{2} + 3\logz{1} - \fr32\logz{3}
%  \biggr) 
%  \biggr] \logT
%  + \nn &  & + 
%  \Bigl[
%    -6(N+8)\pi^2 + 72 (N+2) 
%  \Bigr] \ln \biggl( \frac{g}{\pi} \sqrt{\frac{N+2}{72}} \biggr)
%  + % \nn &  & + 
%  N^2 \biggl[ 6 - 12 \gammaE + \gammaE^2 + \frac{\zeta(3)}{36} \biggr] 
%  + \nn &  & + 
%  16 N \Bigl[ -0.9991160242(2) \Bigr]
%  + % \nn &  & + 
%  64 \Bigl[ -9.0905637831(3) \Bigr]
%  \biggr\}
%  \;. \la{p6}
 p_6 & = & 
 -\frac{5}{108} (N+2) 
 \biggl\{
 \Bigl[
   72 (N+2)  -6(N+8)\pi^2 
 \Bigr] \ln \biggl( \frac{g}{\pi} \sqrt{\frac{N+2}{72}} \biggr)
  + \nn & & +
    (N+8)^2 \biggl( \logT  \biggr)^2  
  + \nn & & +
  N^2 \biggl[ \biggl(2 \gammaE - 12\biggr) \logT + 
  6 - 12 \gammaE + \gammaE^2 + \frac{\zeta(3)}{36} \biggr]  
  + \nn &  & + 
  16 N \biggl[ \biggl(
   -\frac{493}{80} + \frac{5\gammaE}{4} + \fr32\logz{1}-\fr34\logz{3} 
  \biggr) \logT 
  -0.9991160242(2) \biggr]
  + \nn &  & + 
  64 \biggl[ \biggl( 
 -\frac{127}{160} + \frac{\gammaE}{2} + 3\logz{1} - \fr32\logz{3}
 \biggr) \logT
 -9.0905637831(3) \biggr]
 \biggr\}
 \;. \la{p6}
\ea

A number of simple crosschecks can be made on \eqs\nr{complete}--\nr{p6}.
Making use of the RG equation, 
\be
 \bmu \frac{{\rm d} g}{{\rm d}\bmu}
 = \frac{g^3}{(4\pi)^2} \frac{N+8}{6} 
 - \frac{g^5}{(4\pi)^4} \frac{3N+14}{6} 
 + \rmO(g^7)
 \;, \la{RG}
\ee
it is easy to verify that the result is RG
invariant up to the order computed. Setting $N=1$, terms up to $\rmO(g^5)$
agree with Refs.~\cite{az,ps,bn1}. Finally, taking the limit $N\to\infty$
with $g^2 N$ fixed, we get
\ba
  \frac{p(T)}{T^4} & \approx & \frac{\pi^2 N}{90}
  \biggl\{ 
  1  
  + \mathcal{G}^2 % \biggl( \frac{g\sqrt{N}}{4\pi} \biggr)^{\!\! 2}
    \biggl( -\frac{5}{12} \biggr)
  + \mathcal{G}^3 % \biggl( \frac{g\sqrt{N}}{4\pi} \biggr)^{\!\! 3}
    \biggl( \frac{5\sqrt{2}}{9} \biggr)
  + \nn & + & 
    \mathcal{G}^4 % \biggl( \frac{g\sqrt{N}}{4\pi} \biggr)^{\!\! 4}
    \biggl( \frac{5}{36} \biggr)
    \biggl( \logT + \gammaE - 6 \biggr)
  + 
    \mathcal{G}^5 % \biggl( \frac{g\sqrt{N}}{4\pi} \biggr)^{\!\! 5}
    \biggl(- \frac{5}{9\sqrt{2}} \biggr)
    \biggl( \logT + \gammaE - \fr32 \biggr)
  + \nn & + & 
    \mathcal{G}^6 % \biggl( \frac{g\sqrt{N}}{4\pi} \biggr)^{\!\! 6}
    \biggl(- \frac{5}{108} \biggr)
    \biggl[ \biggl( \logT \biggr)^{\!\! 2} 
    + 2 ( \gammaE  - 6) \logT 
    + 6 - 12 \gammaE + \gammaE^2 + \frac{\zeta(3)}{36} \biggr]
  \biggr\}
 \;, \nn \la{largeN}
\ea
where $\mathcal{G} \equiv g \sqrt{N}/4\pi$.
This agrees with \eq(5.8) of Ref.~\cite{Drummond:1997cw}; unfortunately 
most of the non-trivial structures, like logarithms of $g$ or the genuine 
4-loop sum-integrals that we were only able to determine numerically, 
disappear in the large-$N$ limit.

Let us finally evaluate our result numerically.
Though the effect is moderate in practice, we reiterate that
we find it convenient not to use \eq\nr{strict} for the 
numerical evaluation, but the unexpanded expression 
$p_\rmi{hard} + p_\rmi{soft}$, defined as a sum of \eqs\nr{psoft}, \nr{phard}.

If we want to present the numerical results as a function of $T$, 
we first have to insert a value for the renormalized quartic 
coupling $g^2$ as a function of the scale $\bmu$, because
the temperature-dependence emerges in connection with logarithms
related to the running of $g^2$ (cf.\ \eqs\nr{complete}--\nr{p6}). 
Defining $b_1\equiv |\beta_1|/(4\pi)^2$ and 
$b_2 \equiv |\beta_2|/(4\pi)^4$, where $\beta_1, \beta_2$
are from \eq\nr{betas}, as well as 
$\alpha \equiv b_2/b_1$ and 
$t \equiv 2 b_1 \ln(\bmu_\rmi{Landau}/\bmu)$, the 2-loop
RG-equation reads
\be
 \frac{{\rm d} g^2}{{\rm d} t} = -g^4 + \alpha g^6
 \;. 
\ee
This equation can be solved up to a boundary condition.
In formal analogy with QCD, we define the boundary condition
such that 
\be
 \bmu_\rmi{Landau} \equiv \lim_{\bmu\to 0}
 \bmu \Bigl[ b_1 g^2 \Bigr] ^{-b_2/2 b_1^2}
 \exp \Bigl[ \frac{1}{2 b_1 g^2} \Bigr]
 \;. 
 \la{LLandau}
\ee
The solution then reduces to the equation 
\be
 \frac{1}{g^2} 
 + \alpha \ln\biggl( \frac{1}{g^2} - \alpha \biggr) = 
 t + \alpha \ln b_1
 \;, 
\ee
which for small $\bmu$ yields the approximate behaviour
\be
 \frac{1}{g^2} \approx 
 2 b_1 \ln\frac{\bmu_\rmi{Landau}}{\bmu}
 - \frac{b_2}{b_1} \ln\biggl(
 2 \ln\frac{\bmu_\rmi{Landau}}{\bmu}
 \biggr)
 \;. 
\ee

Moreover, following Ref.~\cite{adjoint}, 
we define an ``optimal'' scale according to the 
simultaneous 1-loop 
``fastest apparent convergence'' and ``principal of minimal sensitivity''
point obtained for the effective coupling $\hat g_3^2$, defined in 
\eq\nr{lambda3}: 
\be
 \bmu_\rmi{opt} \equiv 4 \pi e^{-\gammaE} T      
 \;. \la{Lopt}
\ee
The scale $\bmu$ will then be varied in the range
$(0.5...2.0)\bmu_\rmi{opt}$ around this point. 
Note that the scale choice in \eq\nr{Lopt} leads also 
to a formal simplification of the pressure; for instance
the expression in \eq\nr{largeN} obtains the form
\ba
  \frac{p(T)}{T^4} & \approx & \frac{\pi^2 N}{90}
  \biggl\{ 
  1  - \frac{5}{12} 
  \biggl[ 
  \mathcal{G}^2 
  - \frac{4\sqrt{2}}{3}
  \mathcal{G}^3
  + 2 \mathcal{G}^4
  - \sqrt{2} \mathcal{G}^5
  + \biggl(\fr23 + \frac{\zeta(3)}{324} \biggr) \mathcal{G}^6
  \biggr]
  \biggr\}
 \;. \hspace*{0.5cm} \la{largeNsimple}
\ea

%%%%%%%%%%%%%%%%%%%%%%%%%%%%%%%%% FIGURE %%%%%%%%%%%%%%%%%%%%%%%%%%%%%%%%%
\begin{figure}[t]

%\vspace*{-3cm}

\centerline{%
\epsfysize=5.0cm\epsfbox{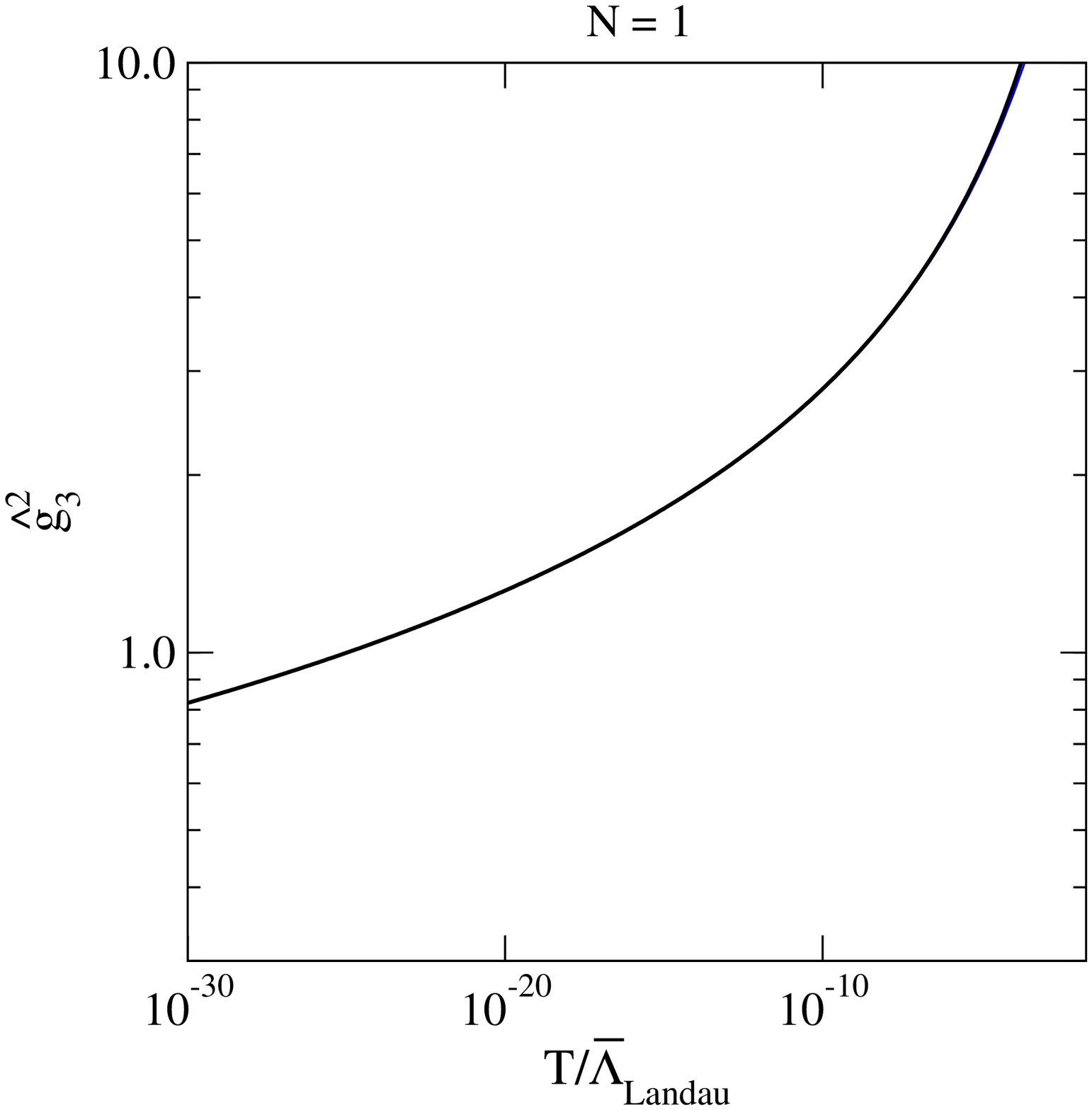}%
~~\epsfysize=5.0cm\epsfbox{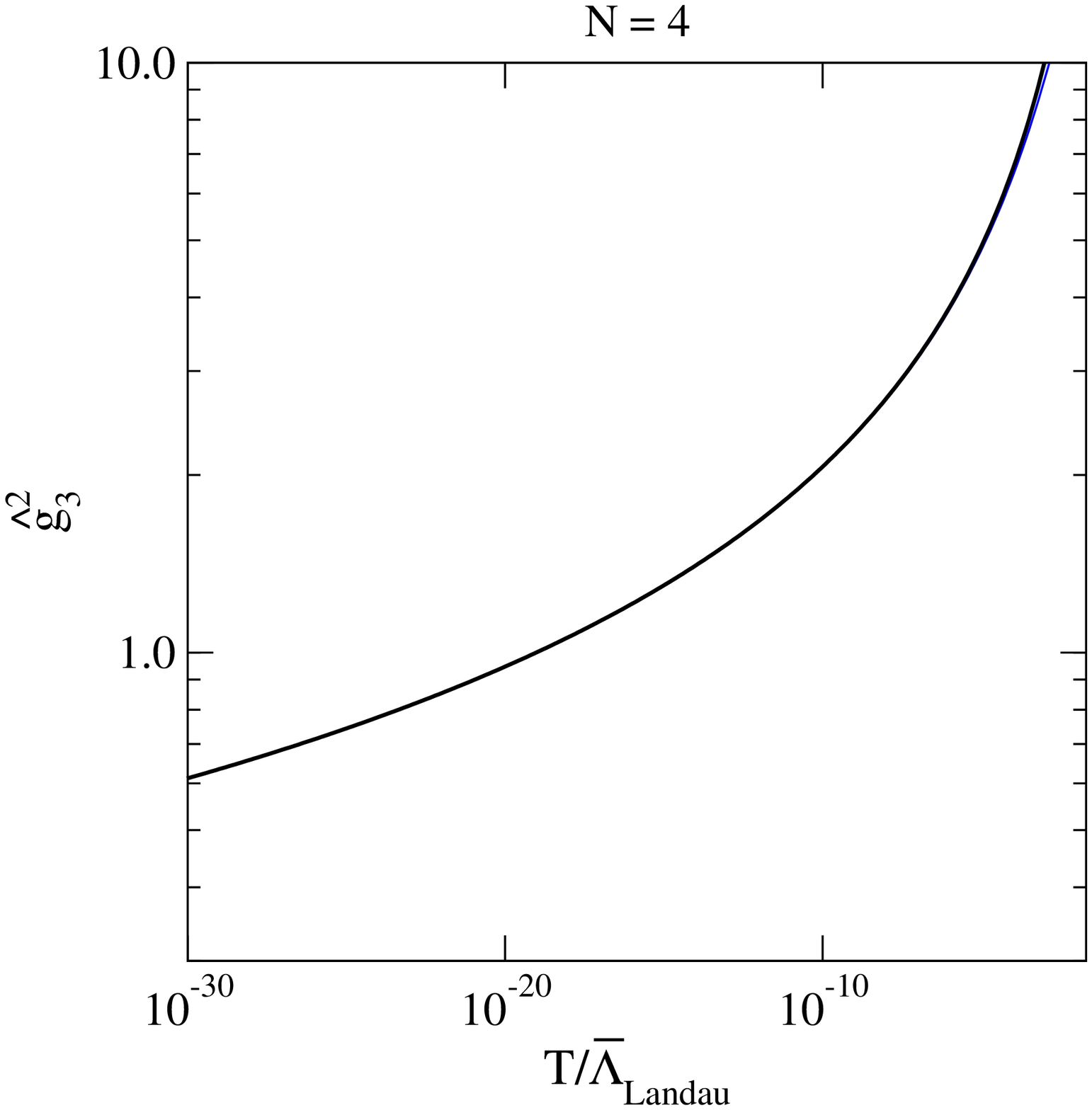}%
~~\epsfysize=5.0cm\epsfbox{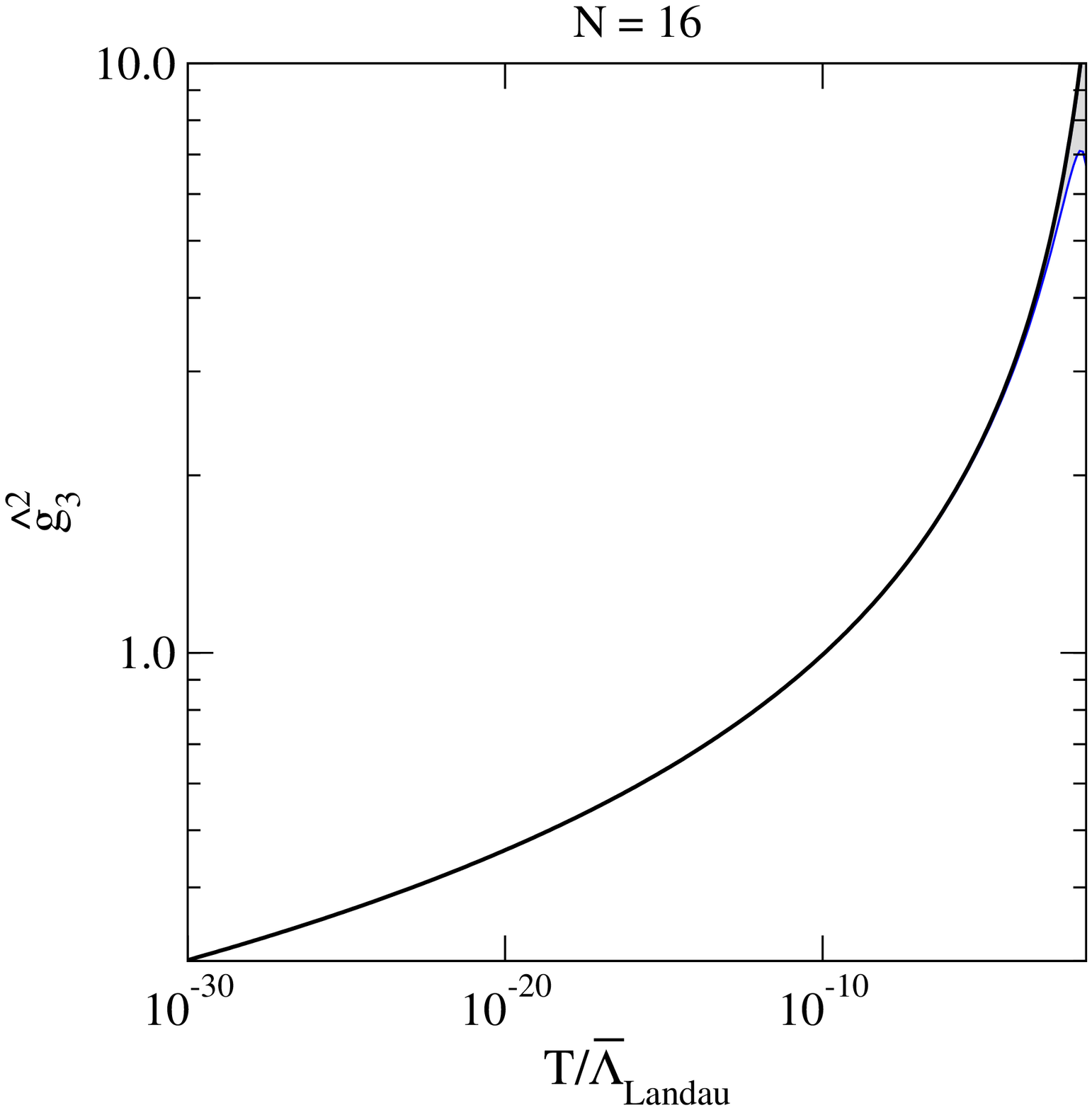}%
}

\caption[a]{\it The effective gauge coupling in \eq\nr{lambda3}, 
as function of $T/\bmu_\rmi{Landau}$, where $\bmu_\rmi{Landau}$
is defined in \eq\nr{LLandau}. From left to right, $N=1,4,16$.
The scale $\bmu$ has been varied within the range 
$(0.5...2.0) \bmu_\rmi{opt}$ (the grey band), with $\bmu_\rmi{opt}$
defined in \eq\nr{Lopt}, but the effects are almost invisible
on our resolution.} 
\la{fig:g32hat}

\end{figure}
%%%%%%%%%%%%%%%%%%%%%%%%%%%%%%%%%%%%%%%%%%%%%%%%%%%%%%%%%%%%%%%%%%%%%%%%%%%

We now first plot $\hat g_3^2$ as a function of $T/\bmu_\rmi{Landau}$
and $N$. The results are shown 
in \fig\ref{fig:g32hat}. The scale $\bmu$ is chosen according
to \eq\nr{Lopt} but the dependence 
on the choice is so small within the range mentioned
that it is almost invisible.

%%%%%%%%%%%%%%%%%%%%%%%%%%%%%%%%% FIGURE %%%%%%%%%%%%%%%%%%%%%%%%%%%%%%%%%
\begin{figure}[t]

%\vspace*{-3cm}

\centerline{%
\epsfysize=5.0cm\epsfbox{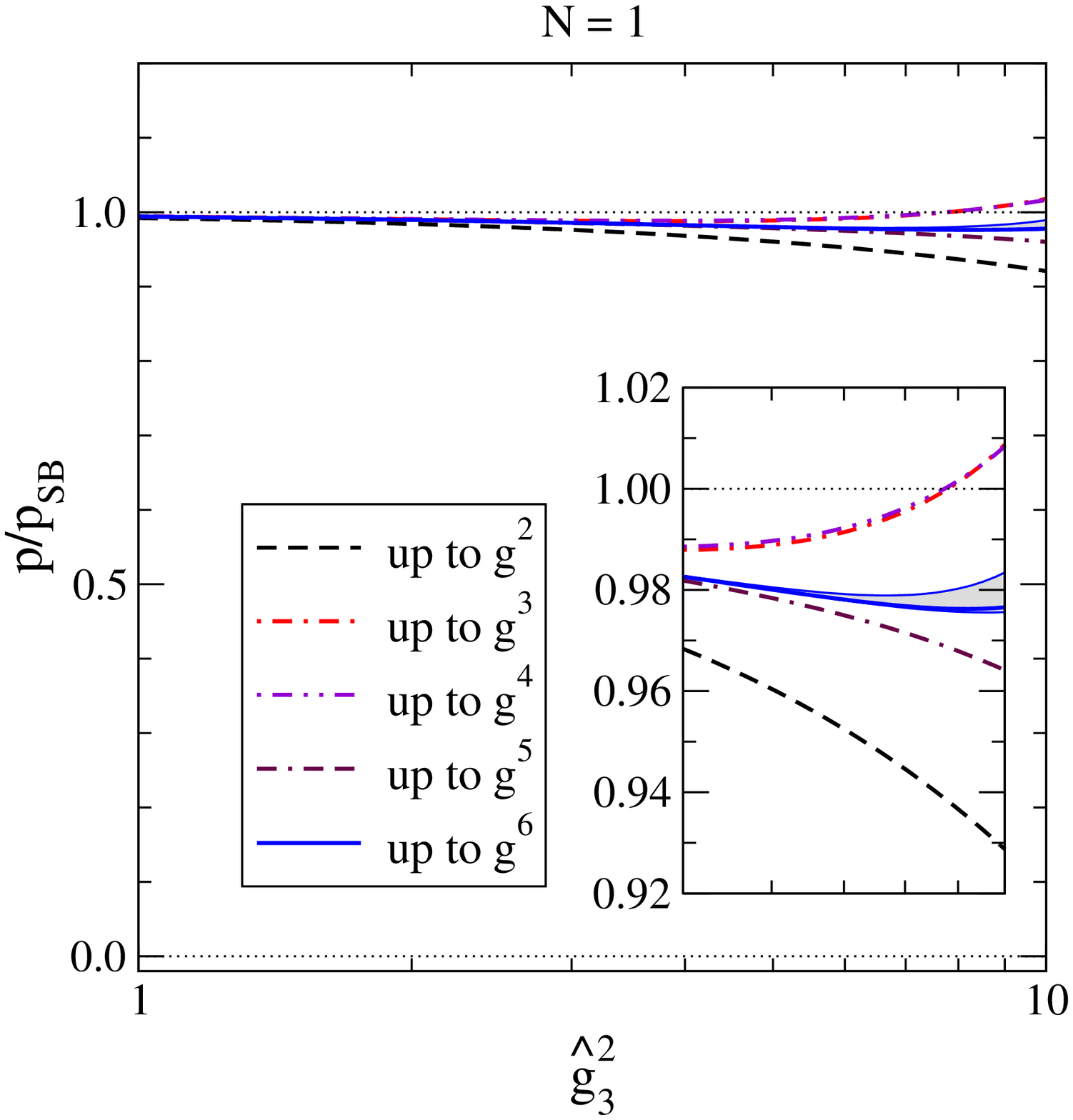}%
~~\epsfysize=5.0cm\epsfbox{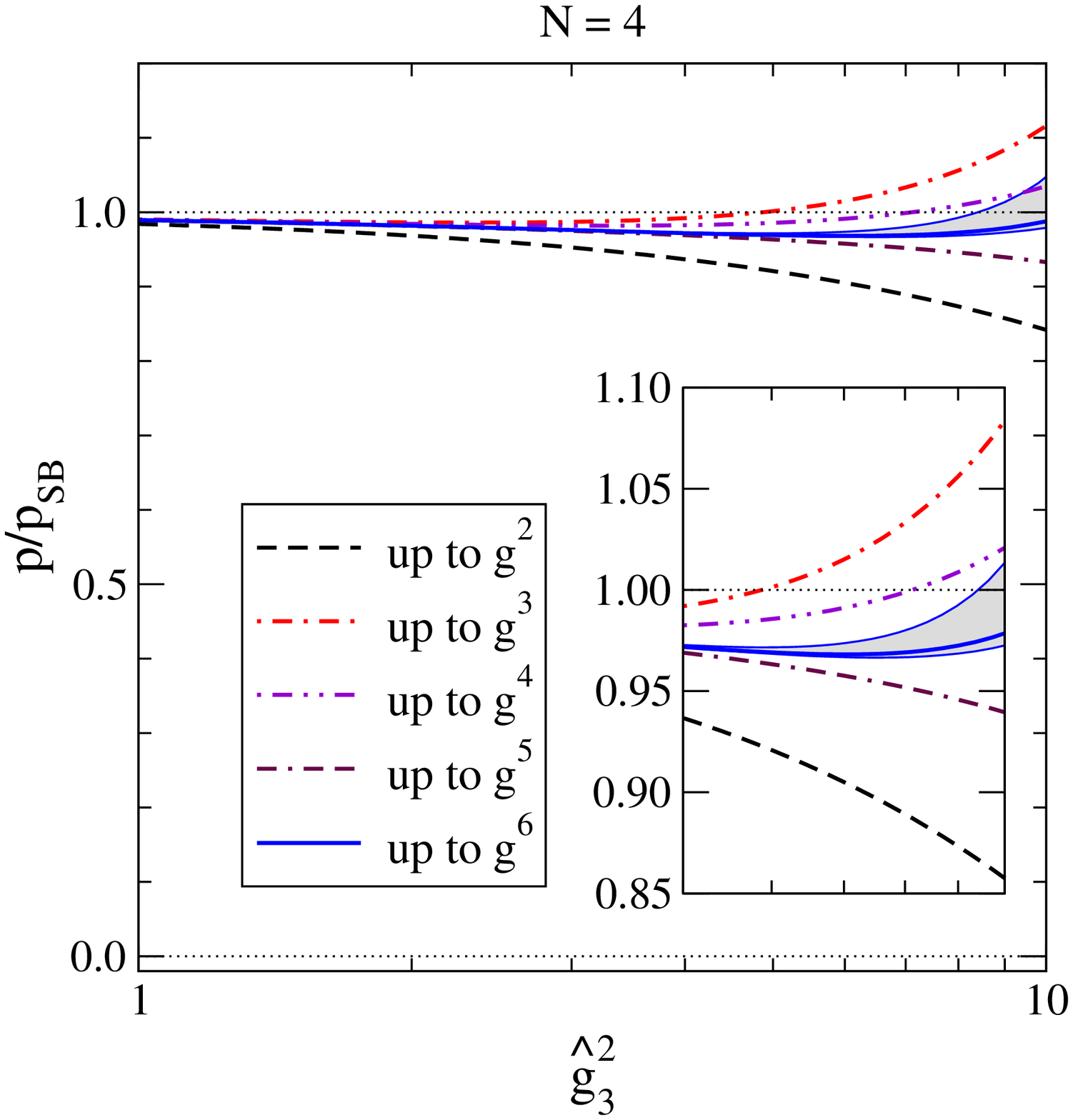}%
~~\epsfysize=5.0cm\epsfbox{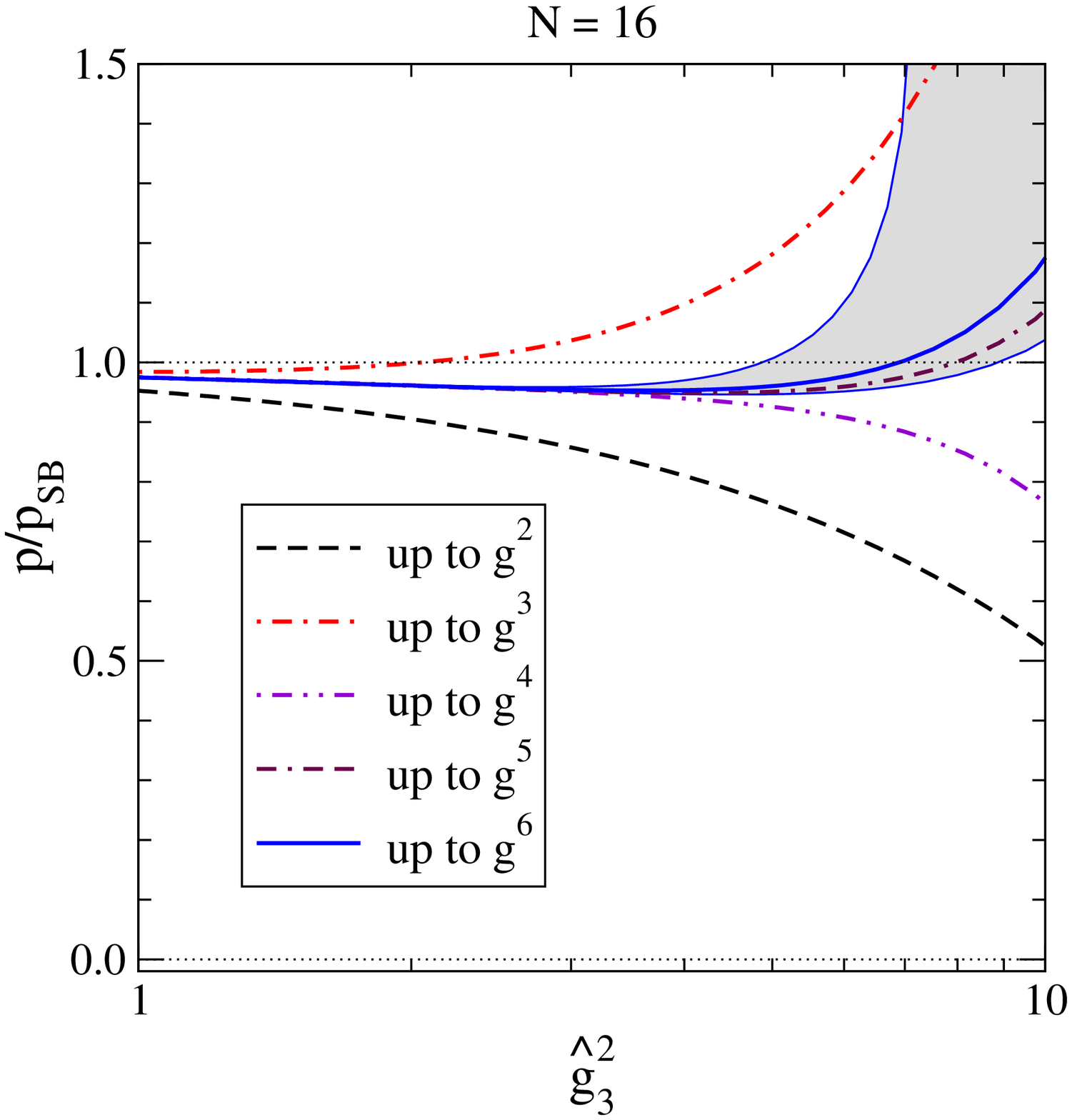}%
}

\caption[a]{\it The resummed perturbative result,
$p(T) = p_\rmi{hard}(T) + p_\rmi{soft}(T)$, normalised
to the free Stefan-Boltzmann result denoted by $p_\rmi{SB}$, 
as a function of the effective gauge coupling shown in \fig\ref{fig:g32hat}.
From left to right, $N=1,4,16$.
For the contribution of $\rmO(g^6)$, 
the scale $\bmu$ has been varied within the range 
$(0.5...2.0) \bmu_\rmi{opt}$ (the grey band), with $\bmu_\rmi{opt}$
defined in \eq\nr{Lopt}.} 
\la{fig:press}

\end{figure}
%%%%%%%%%%%%%%%%%%%%%%%%%%%%%%%%%%%%%%%%%%%%%%%%%%%%%%%%%%%%%%%%%%%%%%%%%%%

Finally we plot the pressure, normalised to the free result, 
in \fig\ref{fig:press}. We have used an overall resolution as would be 
relevant for QCD, where one hopes to reach an accuracy on the 10\% level 
or so; the inserts display finer structures that are invisible
on this resolution but may still be of some academic interest. 
Since $T/\bmu_\rmi{Landau}$ is a quantity
with which it is difficult to associate anything in QCD, we choose
to use $\hat g_3^2$ as the horizontal axis in this figure.  If desired, 
the conversion to $T/\bmu_\rmi{Landau}$ goes through
\fig\ref{fig:g32hat}.

It can be seen that the patterns in \fig\ref{fig:press}
are somewhat similar to 
those familiar from QCD: the order $\rmO(g^3)$ strongly
overshoots the free result, and the subsequent orders then slowly
converge towards a common value. However, for $N=16$, 
{i.e.}\ the case with the same number of scalar degrees of freedom
as QCD, control is certainly lost once $\hat g_3^2 \gg 4.0$.
It is perhaps also worth stressing that even at $N=16$, the 
results deviate fairly substantially from \eq\nr{largeNsimple}; 
relative corrections to the large-$N$ limit can be as large 
as $\sim 8/N$, cf.\ \eq\nr{p6}.

%%%%%%%%%%%%%%%%%%%%%%%%%%%%%%%%% FIGURE %%%%%%%%%%%%%%%%%%%%%%%%%%%%%%%%%
\begin{figure}[t]

%\vspace*{-3cm}

\centerline{%
\epsfysize=7.5cm\epsfbox{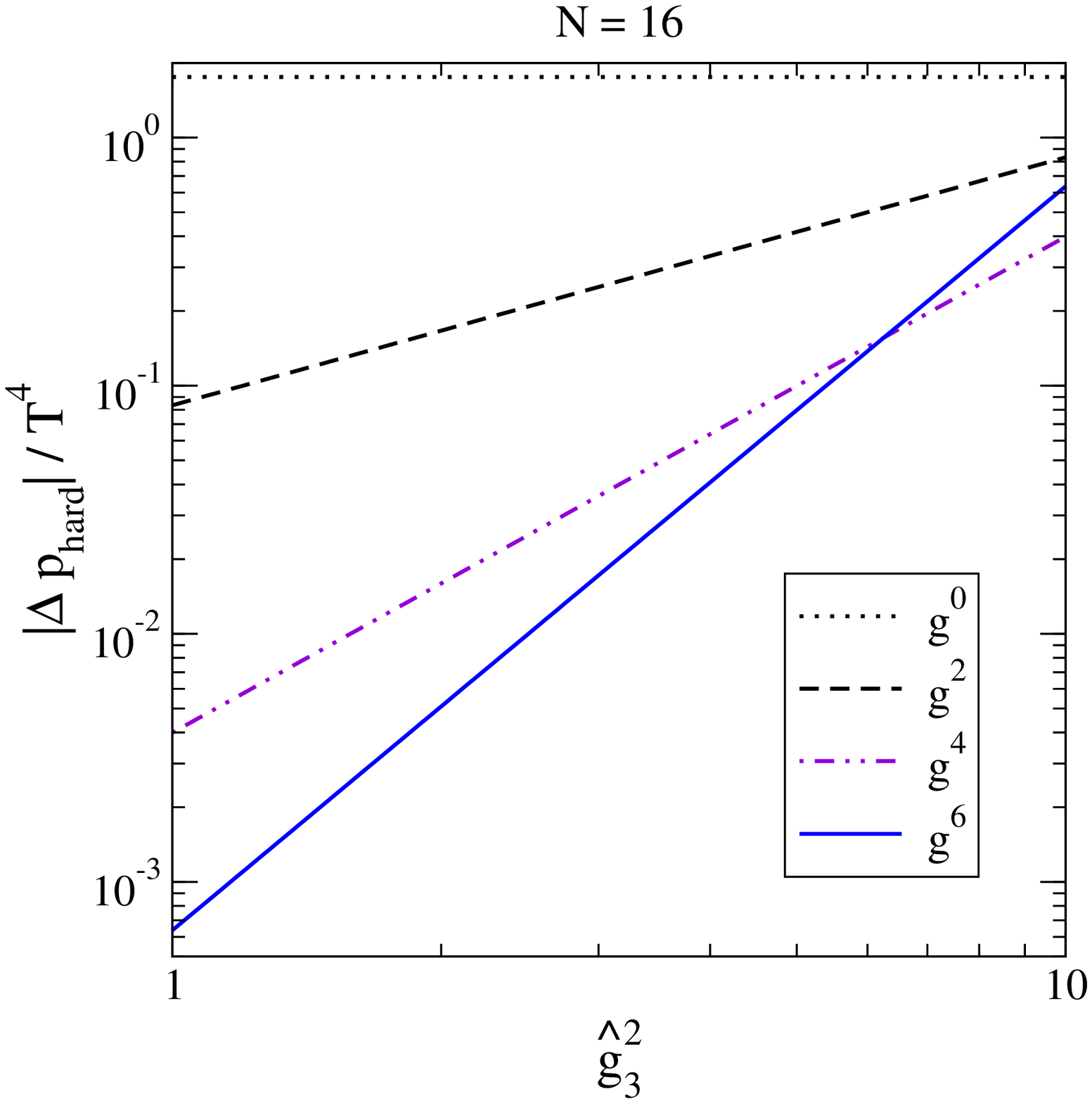}%
~~\epsfysize=7.5cm\epsfbox{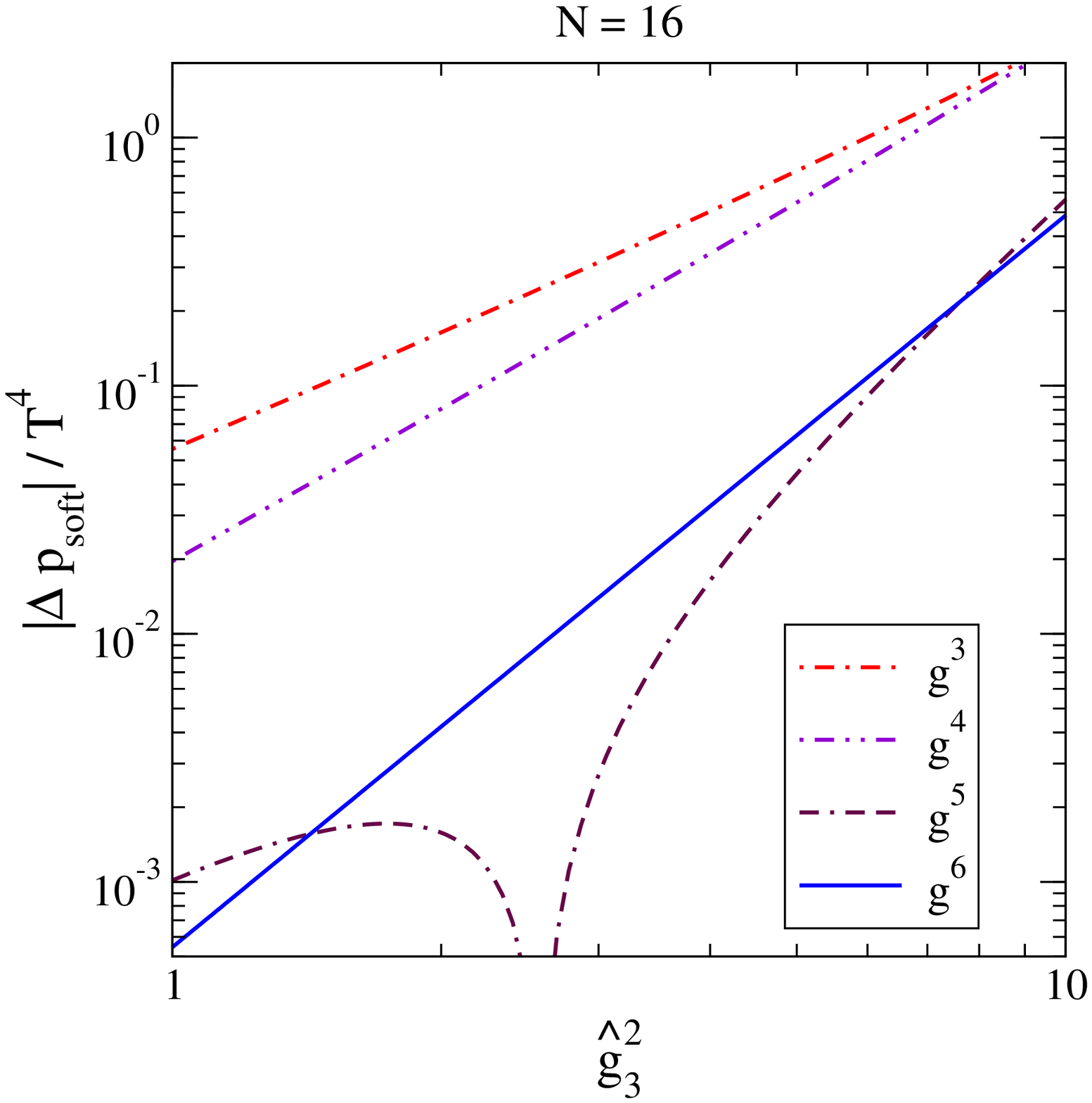}%
}

\caption[a]{\it The absolute values of the various 
order contributions to 
$p_\rmi{hard}/T^4$ (left)
and 
$p_\rmi{soft}/T^4$ (right), 
as a function of the effective gauge coupling shown 
in \fig\ref{fig:g32hat}, for $N=16$.
The convergence appears to be better for $p_\rmi{hard}/T^4$, but
$p_\rmi{soft}/T^4$ shows some convergence as well.  
} 
\la{fig:separ}

\end{figure}
%%%%%%%%%%%%%%%%%%%%%%%%%%%%%%%%%%%%%%%%%%%%%%%%%%%%%%%%%%%%%%%%%%%%%%%%%%%

A perhaps more satisfying view on the result can be obtained
if the hard and the soft contributions to the pressure, 
$p_\rmi{hard}$ and $p_\rmi{soft}$, are plotted separately. 
This has been done in \fig\ref{fig:separ} for $N=16$, 
where the absolute values of each new order are shown. 
It is clear that at least the expansion for $p_\rmi{hard}$ 
does appear to converge up to $\hat g_3^2 \sim 5.0$; 
for $p_\rmi{soft}$ the situation is worse, since
the $\rmO(g^6)$ result is larger than the $\rmO(g^5)$ result
even at fairly small $\hat g_3^2$. This is, however, due to 
the fact that the $\rmO(g^5)$ result crosses zero at
$\hat g_3^2 \approx 2.5$, 
and does not necessarily signal a total breakdown
of the series.

%%%%%%%%%%%%%%%%%%%%%%%%%%% SECTION %%%%%%%%%%%%%%%%%%%%%%%%%%%%%%%%%%%%%
%
\section{Conclusions}
\la{se:concl}

We have computed in this paper the pressure 
of O($N$) scalar field theory up to order $g^6$
in the weak-coupling expansion. In terms of the loop expansion, 
this corresponds to the inclusion of all 4-loop diagrams, 
as well as infinite subsets of higher-loop diagrams 
needed in order to cancel the infrared divergence
of the naive perturbative computation. 

The main motivation for this paper has been ``technological'':
we have demonstrated that 
4-loop sum-integrals are doable, with divergent parts
and logarithms handled analytically, and constant parts
evaluated numerically. The essential ingredients allowing
for the evaluation of the single genuine 4-loop sum-integral
that appeared in our computation,  denoted by $S_2$, were the realization 
that renormalizing the theory before carrying out the sum-integral, allows 
to simplify the structure that needs to be considered (cf.\ \eq\nr{S2def}), 
as well as an application of mixed coordinate and momentum-space
techniques (cf.\ Appendix A.3).

With further work, the constant parts of the 4-loop 
pressure (the numerical values in \eq\nr{p6}) 
might also be doable analytically. This, however, is not 
necessary from the QCD point of view, where the 4-loop contribution 
in any case involves a non-perturbative term~\cite{linde,gpy}
that can only be determined numerically~\cite{plaq,nspt_mass}. 

The complete result, up to $\rmO(g^6)$, 
is shown in \eqs\nr{complete}--\nr{p6}. A more compact form, resumming
hard contributions, is given by a combination 
of \eqs\nr{psoft}, \nr{phard}, with coefficients 
given in \eqs\nr{aE4}--\nr{aE7} and \nr{aE1}--\nr{bE1}.

Though we have thus demonstrated the feasibility
of going to 4-loop level in thermal field theory, 
we would at the same time like to stress 
that the present computation was based on a ``brute force''
approach for the evaluation of the 3-loop and 4-loop 
sum-integrals (cf.\ Appendix A). 
Such an approach requires a lot of patience, 
and is simultaneously susceptible to errors. Thinking about the case
of QCD, where a much larger set of genuine 4-loop integrals will 
appear, it would clearly be most 
desirable to develop somewhat more automatised 
techniques for the evaluation of the loop integrals, in analogy
with what has been achieved in recent years for massive vacuum 
integrals at zero temperature~\cite{laporta,sv}, 
in order to carry out the computation in a more controllable fashion. 

Finally, we should point out that 
the model we have considered has also been 
a popular testing ground for many different theoretical 
tools as well as improved approximation 
schemes~\cite{Bugrii:1995vn}--\cite{Blaizot:2006rj},
in the latter case with the goal of learning something about 
their numerical convergence. We have shown here that, 
for $N=16$,  the resummed weak-coupling expansion 
does appear to show some convergence up to the point where 
the scale-invariant effective coupling constant $\hat g_3^2$, 
defined in \eq\nr{lambda3}, reaches values $\hat g_3^2 \sim 4.0$. 
This includes all values relevant for QCD
above the temperature $T\sim\Lambdamsbar$~\cite{gE2}, 
and therefore should perhaps be interpreted as a positive feature from 
the point of view of the applicability of resummed
perturbation theory at high temperatures, even though it of course
must be stressed that such a direct comparison between
scalar field theory and QCD is far too naive.

\section*{Acknowledgements}

We are grateful to K.~Kajantie for useful discussions. 
A.G. was supported by 
the Natural Sciences and Engineering Research Council of Canada, 
and A.V. % was supported
in part by the U.S.\ Department of Energy under 
Grant No.\ DE-FG02-96ER-40956.
M.L. thanks ECT*, Trento, 
and A.V. thanks Bielefeld University as well as 
the Galileo Galilei Institute and INFN, 
Florence, for hospitality during times when essential 
progress was made with this work. 

% \newpage

%-------------------------------------------------------------------

\appendix
\renewcommand{\thesection}{Appendix~\Alph{section}}
\renewcommand{\thesubsection}{\Alph{section}.\arabic{subsection}}
\renewcommand{\theequation}{\Alph{section}.\arabic{equation}}

%%%%%%%%%%%%%%%%%%%%%%%%%%%%%%%%%%%%%%%%%%%%%%%%%%%%%%%%%%%%%%%%%%%%%%%%

%%%%%%%%%%%%%%%%%%%%%%%%% SECTION %%%%%%%%%%%%%%%%%%%%%%%%%%%%%%%%%%%%%
%
\section{Details of the computation}

%%%%%%%%%%%%%%%%%%%%% SUBSECTION %%%%%%%%%%%%%%%%%%%%%%%%%%%%%%%%%%%%%%%%%
%
\subsection{Self-energies}

%%%%%%%%%%%%%%%%%%%%%%%% SUBSUBSECT %%%%%%%%%%%%%%%%%%%%%%%%%%%%%%%%%%%%%
%
\subsubsection{$\Pi(P)$}

Before attacking the two sum-integrals in \eqs\nr{S1def}, \nr{S2def},  
let us review a few straightforwardly verifiable 
properties of the functions $\Pi(P)$, 
and $\bar{\Pi}(P)$, defined in \eqs\nr{Pi}, \nr{bPi}.
[We will also define a third similar function 
in \se\ref{se:Pitilde}, denoted by $\widetilde{\Pi}(P)$.]  
As has been shown in Ref.~\cite{az}, 
the first of these can be written in the form
\ba
 \Pi(P)&=&\Pi^{(0)}(P)+\fr{2{\cal I}_1}{P^2}+\Delta \Pi(P)
 \;, \label{pi}
\ea
where
\ba
 \Pi^{(0)}(P)&\equiv& \fr{\beta}{(P^2)^{\e}}
 \;\;=\;\;\fr{\Lambda^{2\e}}{(4\pi)^{2-\e}}
 \fr{\Gamma(\e)\Gamma^2(1-\e)}{\Gamma(2-2\e)}\fr{1}{(P^2)^{\e}}
 \la{Pi0}
\ea
denotes its zero-temperature limit and is responsible for 
its leading UV ($P^2\rightarrow \infty$) behaviour. 
The leading UV behaviour of the rest, {i.e.~}the 
finite-temperature part of the function, is on the other hand 
obtainable by letting $P$ become arbitrarily large in either 
of the two propagators of $\Pi(P)$ and then integrating over the
other, which produces $2{\cal I}_1/P^2$. Subsequently, 
the remaining part of the function, denoted here by $\Delta \Pi(P)$, 
behaves in the UV as $1/P^4$.

At $\e=0$, the UV-finite subtracted function $\Delta \Pi(P)$
can be written in a simple form 
by using a three-dimensional (spatial) Fourier transform. 
This gives \cite{az}
\ba
 \Delta\Pi(P)&=&\fr{T}{(4\pi)^2}\int\!{\rm d}^3\vec{r} 
  \fr{1}{r^2}e^{i\mathbf{p}\cdot\mathbf{r}} e^{-|p_0|r}
 \(\coth \bar{r}-\fr{1}{\bar{r}}-\fr{\bar{r}}{3}\)
 \;, \label{deltapi}
\ea
where $\bar{r}\equiv 2\pi T r$.

Sometimes we need to refer to the finite-temperature part 
of $\Pi(P)$,  defined as $\Pi^{(T)}(P)\equiv \Pi(P)-\Pi^{(0)}(P)$. 
Its Fourier representation after setting $\epsilon\to 0$ is obtained
from Eq.~(\ref{deltapi}) by leaving out the last  term $-\bar{r}/3$ 
from inside the parentheses. In one instance, 
we will also require another version of $\Delta \Pi (P)$, 
in which its leading UV behaviour is further subtracted. 
This function is defined as
\ba
 \Delta' \Pi(P) &\equiv& \Delta\Pi(P)
 -\fr{8T^4J_1}{3-2\e}\(\fr{1}{P^4}-(4-2\e)\fr{p_0^2}{P^6}\)
 \(1-\delta_{p_0}\)
 \;,
\ea
with the Kronecker-$\delta$ ($\delta_{p_0} \equiv \delta_{p_0,0}$)
introduced in the last term 
for convenience. Its Fourier representation  after $\epsilon\to 0$
is equivalent to Eq.~(\ref{deltapi}) apart from having an extra term 
of the form $+\(1-\delta_{p_0}\)\bar{r}^3/45$ inside the parentheses. 
The constant $J_1$ is evaluated in Ref.~\cite{az} and reads
\ba
 J_1 &=& 2^{-2+2\e}\pi^{-3/2+\e}
 \(\fr{\Lambda^2}{T^2}\)^{\!\!\e}
 \fr{\Gamma(4-2\e)}{\Gamma(3/2-\e)}\zeta(4-2\e)
 \;.
\ea

%%%%%%%%%%%%%%%%%%%%%%%% SUBSUBSECT %%%%%%%%%%%%%%%%%%%%%%%%%%%%%%%%%%%%%
%
\subsubsection{$\widetilde{\Pi}(P)$}
\la{se:Pitilde}

It will be convenient in the following to add a third function
to those given in \eqs\nr{Pi}, \nr{bPi}. We define this function through 
\ba
 \widetilde{\Pi}(P)&\equiv&\sumint_Q \fr{\Pi^{(0)}(Q)}{(Q-P)^2}
 \;, \label{Pitilde}
\ea
where $\Pi^{(0)}(Q)$ denotes the zero-temperature part of $\Pi(Q)$
(cf.\ \eq\nr{Pi0}). 

An analogous reasoning as for
$\Pi(P)$ produces for $\widetilde{\Pi}(P)$ the representation
\ba
 \widetilde{\Pi}(P)&=&\widetilde{\Pi}^{(0)}(P)
 +{\cal I}_1\Pi^{(0)}(P)+\Delta \widetilde{\Pi}(P)
 \;, \label{pitilde}
\ea
where
\ba
 \widetilde{\Pi}^{(0)}(P)&\equiv& {\widetilde \beta}
 \(P^2\)^{1-2\e}
 \;\;=\;\;\fr{\Lambda^{2\e}}{(4\pi)^{2-\e}}
 \fr{\Gamma(1-\e)\Gamma(2-2\e)\Gamma(-1+2\e)}
 {\Gamma(\e)\Gamma(3-3\e)}\beta \(P^2\)^{1-2\e}
 \;, \la{tPi0}
\ea
and $\Delta \widetilde{\Pi}(P)$ behaves in the UV
like $1/P^2$. The only difference with respect to the
previous calculation is that when taking the UV limit
of the finite-temperature part of $\widetilde{\Pi}(P)$,
the two ``propagators'' of which $\widetilde{\Pi}(P)$
is composed are not symmetric, and the dominant
$P^2\to\infty$ behaviour is obtained when the large
momentum is routed solely  through the function
$\Pi^{(0)}(Q)$, while the argument of the $1/(Q-P)^2$
propagator is integrated over.

As with $\Pi(P)$, we next derive a spatial Fourier
representation for $\Delta \widetilde{\Pi}(P)$. We will carry out 
the Fourier-transforms strictly in three dimensions, even though
the functions transformed may contain $\e\neq 0$; this is sufficient 
since, as we will see, no divergences in $\e$ appear in the formal
coordinate-space representations (there are in fact divergences 
hidden in the lower ranges of the $\vec{r}$-integration in 
individual terms, but they cancel for the finite quantity 
$\Delta \widetilde{\Pi}(P)$ that we are ultimately interested in). 
Defining the (inverse) Fourier transform
\be
 \mathcal{F}(q_0,r;\alpha) \equiv  
 \int\!\fr{{\rm d}^3\vec{q}}{(2\pi)^3}
 \fr{e^{-i\mathbf{q}\cdot\mathbf{r}}}{(q^2+q_0^2)^{\alpha}}
 =
 \fr{2^{-1/2-\alpha}\pi^{-3/2}}
 {\Gamma(\alpha)} \(\fr{|q_0|}{r}\)^{3/2-\alpha}
 {\rm K}_{3/2-\alpha}(|q_0|r)
 \;,\label{epsint}
\ee
and making use of \eqs\nr{Pi0}, \nr{tPi0}, 
we obtain the Fourier-representations 
\ba
 \frac{1}{(Q-P)^2} & = & 
 \int\! {\rm d}^3 \vec{r} \, 
 e^{i (\vec{q} - \vec{p})\cdot \vec{r} }
 \, \mathcal{F}(q_0-p_0,r;1)
 \nn  & = & 
 \int\! {\rm d}^3 \vec{r} \, 
 e^{i (\vec{p} - \vec{q})\cdot \vec{r} }
 e^{-|q_0 - p_0| r} \frac{1}{4\pi r}
 \;, \la{prop} \\
%%%
 \Pi^{(0)}(Q) & \approx &   
 \int\! {\rm d}^3 \vec{r} \, 
 e^{i \vec{q}\cdot \vec{r} }
 \frac{\Lambda^{2\e}}{(4\pi)^{2-\e}}  
 \fr{ \Gamma^2(1-\e)\Gamma(\e)}{\Gamma(2-2\e)} 
 \; \mathcal{F}(q_0,r;\e) 
 \nn & \approx & 
 \int\! {\rm d}^3 \vec{r} \, 
 e^{i \vec{q}\cdot \vec{r} }
 \biggl\{ 
 e^{-|q_0| r}
 \frac{2}{(4\pi r)^3}
 (1 + |q_0| r) 
 + \rmO(\e) \biggr\}
 \;, \la{pi0fourier} \\ 
%%%
 \widetilde \Pi^{(0)}(Q) & \approx &   
 \int\! {\rm d}^3 \vec{r} \, 
 e^{i \vec{q}\cdot \vec{r} }
 \frac{\Lambda^{4\e}}{(4\pi)^{4-2\e}}  
 \fr{ \Gamma^3(1-\e) \Gamma(-1+2\e)}{\Gamma(3-3\e)} 
 \; \mathcal{F}(q_0,r;-1+2\e) 
 \nn & \approx & 
 \int\! {\rm d}^3 \vec{r} \, 
 e^{i \vec{q}\cdot \vec{r} } 
 \biggl\{ 
 e^{-|q_0| r}
 \frac{2}{(4\pi r)^5}
 (3+ 3|q_0| r+q_0^2 r^2) 
 + \rmO(\e)
 \biggr\}
 \;, \la{tpi0fourier}
\ea
where the symbol ``$\approx$'' is a reminder of the 
fact that the integrals are not well-defined around the origin, 
and we note that the divergent factors 
$\Gamma(\e)$ and $\Gamma(-1+2\e)$ have cancelled 
against the corresponding ones in \eq\nr{epsint}. 
%% The ${\mathcal O}(\e)$ errors on the first lines
%% of \eqs\nr{prop}--\nr{tpi0fourier}
%% come from setting $\e=0$ in the integration measures. 

Inserting now \eqs\nr{prop}, \nr{pi0fourier} into 
the definition of $\widetilde{\Pi}(P)$ in \eq\nr{Pitilde},
produces
\ba
 \widetilde{\Pi}(P)&\approx&
 \fr{2T}{(4\pi)^4}\int\!{\rm d}^3\vec{r}  \,
 e^{i\mathbf{p}\cdot\mathbf{r}}
 \biggl\{ 
 \fr{1}{r^4}
  \sum_{q_0}\(1+|q_0|r\)e^{-|q_0|r}e^{-|q_0-p_0|r} + \rmO(\e) \biggr\}
 \nn &\approx&
 \fr{T}{(4\pi)^4}\int\!{\rm d}^3\vec{r} \,
 e^{i\mathbf{p}\cdot\mathbf{r}}
 \biggl\{  
 e^{-|p_0|r}
 \fr{1}{r^4}
 \biggl[ \bar{r}{\rm csch}^2\bar{r}
 +(2+|\bar{p}_0|\bar{r})(|\bar{p}_0|+\coth\bar{r})\biggr]
 + \rmO(\e) \biggr\}
 \;, \nn \la{pitildefourier}
\ea
with $\bar{p}_0 \equiv {p_0}/{2\pi T}$. Subtracting \eq\nr{tpi0fourier} 
as well as a result obtained from \eq\nr{pi0fourier}, 
\ba
  {\cal I}_1\Pi^{(0)}(P) &\approx& \fr{T}{(4\pi)^4}
 \int\! {\rm d}^3 \vec{r} \, 
 e^{i \vec{p}\cdot \vec{r} }
 \biggl\{ 
 e^{-|p_0| r}
 \frac{1}{r^4} \frac{\bar r}{3} 
 (1 + |p_0| r)
 + \rmO(\e) \biggr\} 
 \;,
\ea
leads to the well-defined form
\ba
 \left. \Delta \widetilde{\Pi}(P) \right|_{\epsilon = 0} &=&
 \fr{T}{(4\pi)^4}\int\!{\rm d}^3\vec{r} \, 
 \fr{1}{r^4}e^{i\mathbf{p}\cdot\mathbf{r}} e^{-|p_0|r}
 \bigg\{\bar{r}{\rm csch}^2\bar{r}
 +(2+|\bar{p}_0|\bar{r})(|\bar{p}_0|+\coth\bar{r})
 - \nn &-&
 \fr{1}{\bar{r}}(3+3|\bar{p}_0|\bar{r}+\bar{p}_0^2\bar{r}^2)
 -\fr{\bar{r}}{3}(1+|\bar{p}_0|\bar{r})\bigg\}
 \;, \label{deltapitilde}
\ea
where the integration is convergent around the origin
such that we have replaced the symbol ``$\approx$'' with equality. 
We note that the expression
inside the curly brackets in \eq\nr{deltapitilde}
behaves at small $\bar{r}$ as $\bar{r}^3$,
and $\Delta \widetilde{\Pi}(P)$ consequently
indeed vanishes at zero temperature, and behaves
as $1/P^2$ at large $P$.

%%%%%%%%%%%%%%%%%%%%%%%% SUBSUBSECT %%%%%%%%%%%%%%%%%%%%%%%%%%%%%%%%%%%%%
%
\subsubsection{$\bar{\Pi}(P)$}

Finally, we divide the function $\bar{\Pi}(P)$ into two parts by 
separating from it the contribution of the Matsubara zero-mode, 
\ba
 \bar{\Pi}(P)&\equiv&\sumint_Q \fr{1}{Q^4(Q-P)^2}
  \;\;=\;\;\sumint '_Q \fr{1}{Q^4(Q-P)^2}+T\Lambda^{2\e}\!\!\int\!
 \fr{{\rm d}^{3-2\e} \vec{q}}{(2\pi)^{3-2\e}}
 \fr{1}{q^4 [(\mathbf{q}-\mathbf{p})^2+p_0^2]}\nn
 &\equiv&\bar{\Pi}_r(P)+\bar{\Pi}_0(P)
 \;,
\ea
where the prime in the upper right corner of 
the sum-integral symbol signifies the leaving out of 
the zero-mode $q_0=0$ from the corresponding Matsubara sum.
Both of these parts are UV-finite. 
In three dimensions, one obtains the representations
\ba
 \bar{\Pi}_r(P)&=&\fr{T}{2(4\pi)^2}\!\int\!{\rm d}^3\vec{r} \,
 \fr{1}{r} e^{i\mathbf{p}\cdot\mathbf{r}}
 \sum_{q_0\neq 0}\fr{1}{|q_0|}e^{-(|q_0|+|q_0-p_0|)r}
 \;, \label{pibar}\\
 \bar{\Pi}_0(P)&=&-\fr{T}{(2\pi)^2}\int_0^{\infty}\!\!{\rm d}q
 \fr{1}{q^2}\bigg\{\fr{2}{P^2}
  +\fr{1}{2pq}\ln\bigg[\fr{(p-q)^2+p_0^2}{(p+q)^2+p_0^2}\bigg]\bigg\}
 \nn &=&-\fr{T}{4\pi}\fr{|p_0|}{P^4} 
 \;\;=\;\;-\fr{T}{2(4\pi)^2}\!\int\!{\rm d}^3\vec{r} \,
  e^{i\mathbf{p}\cdot\mathbf{r}}e^{-|p_0|r}
 \;, \label{pibar0}
\ea
where in the latter case we have first (in $d=3-2\e$ dimensions) 
subtracted from the integrand the term ${1} / {q^4P^2}$, 
the integral of which vanishes in dimensional regularization 
but which renders the expression for $\bar{\Pi}_0(P)$ IR convergent.

%%%%%%%%%%%%%%%%%%%%%%%%%% SUBSECTION %%%%%%%%%%%%%%%%%%%%%%%%%%%%%%%%%%%%%
%
\subsection{Strategy for determining $S_1$}

As often in the evaluation of multi-loop (sum-)integrals, the first and 
in some sense also the most important step is to divide the integrand into 
two types of terms: ones that are divergent, but sufficiently simple 
to allow for an analytic evaluation, and others that are perhaps 
complicated but both IR and UV convergent.
In the case of the sum-integral $S_1$ defined in \eq\nr{S1def}, 
we do this by first decomposing $\Pi(P)$ 
according to Eq.~(\ref{pi}), and then interchanging the order 
of sum-integrals, $P\leftrightarrow Q$, in the terms involving
$\Pi^{(0)}(P)$ and $2 \mathcal{I}_1/P^2$. 
This leads to the following decomposition 
of the original sum-integral:
\ba
 S_1&=&\sumint_P \Delta \Pi(P)\bar{\Pi}_r(P)
     +\sumint_P \Delta \Pi(P)\bar{\Pi}_0(P) 
     +2{\cal I}_1\sumint_P\fr{\Pi(P)}{P^4}
     +\sumint_P\fr{\widetilde{\Pi}(P)}{P^4}
 \;, \nn
 &\equiv& S_1^\rmi{I} + S_1^\rmi{II} + S_1^\rmi{III} + S_1^\rmi{IV}
 \;.
\ea
In $S_1^\rmi{IV}$ we repeat the split-up 
procedure by using Eq.~(\ref{pitilde}) for $\widetilde{\Pi}(P)$. 
Furthermore, we take care of  
IR divergences by separating the contribution 
of the zero-mode from the Matsubara sums where necessary: 
\ba
 S_1^\rmi{II} &=& 
 \sumint '_P \Delta \Pi(P)\bar{\Pi}_0(P)+ 
 T\Lambda^{2\e}\!\!\int\! \fr{{\rm d}^{3-2\e} \vec{p}}{(2\pi)^{3-2\e}}
 \Pi(p_0=0,p)\bar{\Pi}_0(p_0=0,p)\nn
 &\equiv& S_1^\rmi{II,a} + S_1^\rmi{II,b}
 \;, \\
 S_1^\rmi{III}&= & 2 {\cal I}_1 \Bigg\{
 \sumint '_P \fr{\Pi^{(T)}(P)}{P^4} 
 + T\Lambda^{2\e}\!\!\int\! \fr{{\rm d}^{3-2\e} \vec{p}}{(2\pi)^{3-2\e}}
 \fr{\Pi(p_0=0,p)}{p^4} +
 \sumint_P \fr{\Pi^{(0)}(P)}{P^4}\Bigg\}
 \nn
 &\equiv& S_1^\rmi{III,a} + S_1^\rmi{III,b} + S_1^\rmi{III,c}
 \;,\\
 S_1^\rmi{IV}&=&\sumint '_P\fr{\Delta\widetilde{\Pi}(P)}{P^4} 
 + T\Lambda^{2\e}\!\!\int\! \fr{{\rm d}^{3-2\e} \vec{p}}{(2\pi)^{3-2\e}}
 \fr{\widetilde{\Pi}(p_0=0,p)}{p^4}
 +\beta {\cal I}_1\sumint_P \fr{1}{(P^2)^{2+\e}}
 +\sumint_P \fr{\widetilde{\Pi}^{(0)}(P)}{P^4}
 \nn
 &\equiv& S_1^\rmi{IV,a} + S_1^\rmi{IV,b} + S_1^\rmi{IV,c} + S_1^\rmi{IV,d} 
 \;.
\ea
The only further manipulations performed here are the dropping 
of terms that vanish in dimensional regularization; 
this happens in Matsubara zero-mode integrals
without scales, after writing 
$
 \Delta\Pi \to \Pi - \Pi^{(0)} - 2 \mathcal{I}_1/P^2 
$ 
in $S_1^\rmi{II,b}$, 
$ 
 \Pi^{(T)}\to \Pi - \Pi^{(0)}
$ 
in $S_1^\rmi{III,b}$, and 
$
 \Delta\widetilde{\Pi} \to \widetilde{\Pi} -
 \widetilde{\Pi}^{(0)} - \mathcal{I}_1 \Pi^{(0)} 
$
in $S_1^\rmi{IV,b}$.

We have now separated the original sum-integral into ten pieces, 
which % --- as we will see --- 
fall into the following sub-categories:
\begin{itemize}
\item Finite terms that can be evaluated numerically: 
      $S_1^\rmi{I}$, $S_1^\rmi{II,a}$, $S_1^\rmi{III,a}$, $S_1^\rmi{IV,a}$.
\item Possibly divergent terms that can be evaluated analytically through 
      the introduction of Feynman parameters: 
      $S_1^\rmi{II,b}$, $S_1^\rmi{III,b}$, $S_1^\rmi{IV,b}$.
\item Terms that are trivial to compute due to Eq.~(\ref{in}): 
      $S_1^\rmi{III,c}$, $S_1^\rmi{IV,c}$, $S_1^\rmi{IV,d}$.
\end{itemize}
Below, we will go through the evaluation of the sum-integrals 
of the first category in detail and outline the calculation 
of those in the second category.

\subsubsection{$S_1^\rmi{I}$}

It is straightforward to see that $S_1^\rmi{I}$ is finite 
in three dimensions, so we will immediately set $\e=0$. 
Expressing $\Delta \Pi(P)$ and $\bar{\Pi}_r(P)$ in terms of
their Fourier representations, given in Eqs.~(\ref{deltapi}) 
and (\ref{pibar}), enables us to perform the $\vec{p}$-integral 
in the definition of $S_1^\rmi{I}$ to give a 
$\delta$-function in coordinate space. 
This leaves us with the result
\ba
 S_1^\rmi{I} &=&\fr{T^3}{2(4\pi)^4}\int\!{\rm d}^3\vec{r} 
  \fr{1}{r^3}\(\coth \bar{r}-\fr{1}{\bar{r}}-\fr{\bar{r}}{3}\)
  \sum_{p_0}\sum_{q_0\neq 0}
 \fr{e^{-(|p_0|+|q_0|+|q_0-p_0|)r}}{|q_0|}
 \nn &=&
 \fr{2T^2}{(4\pi)^4}\int_0^{\infty}\!\!{\rm d}r\,\fr{1}{r}
  \(\coth r-\fr{1}{r}-\fr{r}{3}\)\bigg\{\fr{1}{e^{2r}-1}
 -\coth r\, \ln(1-e^{-2r})\bigg\}\nn
 &\approx&-\fr{T^2}{(4\pi)^4}\times 
 0.0269726622737(1) \;, 
\ea
where we have analytically performed 
the sums over $p_0$ and $q_0$ (in this order) and later dropped 
the bars over the dimensionless coordinate variable.
The last one-dimensional integral is of a form that might allow for 
an analytic evaluation, but for the purposes 
of this paper we have simply computed its value numerically.
The number in parentheses estimates the uncertainty of the last digit.

\subsubsection{$S_1^\rmi{II,a}$}

For $S_1^\rmi{II,a}$, we proceed along the lines of the previous section, 
employing this time the Fourier representation of $\bar{\Pi}_0(P)$ given 
in Eq.~(\ref{pibar0}). After scaling all variables dimensionless, we obtain
\ba
 S_1^\rmi{II,a}&=&-\fr{T^2}{(4\pi)^5}\int\!{\rm d}^3\vec{r} 
  \fr{1}{r^2}\(\coth r-\fr{1}{r}-\fr{r}{3}\)\sum_{n\neq 0}e^{-2|n|r} 
  \nn &=&
 -\fr{2T^2}{(4\pi)^4}
 \int_0^{\infty}\!\!{\rm d}r\(\coth r-\fr{1}{r}-\fr{r}{3}\)\fr{1}{e^{2r}-1}
 \nn &\approx& \fr{T^2}{(4\pi)^4}\times 0.0134942763002(1)
 \;,
\ea
where the last integration was performed numerically.\footnote{%
 Its analytic value is 
 $T^2 [1 + \gammaE + \pi^2/36 - \ln(2\pi)] /(4\pi)^4$.
 }

\subsubsection{$S_1^\rmi{III,a}$}

Expressing $1/P^4$ and $\Pi^{(T)}(P)$ in terms of their respective
Fourier integrals, we again perform the three-dimensional momentum 
integral and end up with one coordinate space integral and an 
infinite Matsubara sum. In dimensionless variables, we get
\ba
 S_1^\rmi{III,a}&=&\fr{4{\cal I}_1}{(4\pi)^5}\int{\rm d}^3\vec{r}\,
 \fr{1}{r^2}\bigg(\coth r - \fr{1}{r}\bigg)
 \sum_{n\neq 0}\fr{1}{|n|}e^{-2|n|r}\nn
 &=&-\fr{2T^2}{3(4\pi)^4}
 \int_0^{\infty}\!\!{\rm d}r\,\bigg(\coth r - \fr{1}{r}\bigg)\ln(1-e^{-2r})
 \nn &\approx&
 \fr{T^2}{(4\pi)^4}\times 0.0625154109468(1)
 \;,
\ea
where the last integration was performed numerically.\footnote{%
 Its analytic value is ${T^2}({\gammaE^2}+{2\gamma_1})/{3}{(4\pi)^4}$,
 where $\gamma_1$ refers to the first Stieltjes gamma constant
 (cf.\ the explanation following \eq\nr{bE3}).
 }

\subsubsection{$S_1^\rmi{IV,a}$}

To obtain $S_1^\rmi{IV,a}$, we repeat the calculation 
of $S_1^\rmi{III,a}$, but simply replace $\Pi^{(T)}(P)$ 
by $\Delta\widetilde{\Pi}(P)$ and accordingly Eq.~(\ref{deltapi})
[without $-\bar r/3$]
by Eq.~(\ref{deltapitilde}). This gives 
\ba
 S_1^\rmi{IV,a}&=&\fr{T^2}{2(4\pi)^5}
 \int{\rm d}^3\vec{r}\,\fr{1}{r^4}
 \sum_{n\neq 0}\fr{1}{|n|}\bigg\{r{\rm csch}^2r
 +(2+|n|r)(|n|+\coth r)
 - \nn & & - 
 \fr{1}{r}(3+3|n|r+n^2r^2)-\fr{r}{3}(1+|n|r)\bigg\}e^{-2|n|r}
 \nn &=&
 \fr{T^2}{(4\pi)^4}\int_0^{\infty}\!\!{\rm d}r\,
 \fr{1}{r^2}\bigg\{\fr{1}{e^{2r}-1}\(r\coth r-1-\fr{r^2}{3}\)
 + \nn & & + 
 \fr{1}{r}\ln(1-e^{-2r})
 \bigg(3+\fr{r^2}{3}-r^2{\rm csch}^2r -2r\coth r\bigg)\bigg\}
 \nn &\approx&
 -\fr{T^2}{(4\pi)^4}\times 0.0004627085472(1)
 \;.
\ea

\subsubsection{$S_1^\rmi{II,b}$, $S_1^\rmi{III,b}$ and $S_1^\rmi{IV,b}$}

The sum-integrals $S_1^\rmi{II,b}$, $S_1^\rmi{III,b}$ and $S_1^\rmi{IV,b}$ 
have one thing in common: the ``outer'' sum-integral is restricted to the
Matsubara zero-mode $p_0 = 0$, while the ``inner'' loop 
contains only one Matsubara sum. This implies that by introducing
Feynman parameters, it is always possible to perform 
all of the momentum integrals analytically, 
leaving in the end a simple sum of a form that
produces just a Riemann $\zeta$-function. As a first step, 
we quote the following results for the ``static limits'' 
($p_0 = 0$) of the various functions, obtained through Feynman
parametrization:
\ba
 \Pi(p_0=0,p)&=&T\Lambda^{2\e}\fr{\Gamma(1/2+\e)}{(4\pi)^{3/2-\e}}
 \sum_{q_0}\int_0^1\!\!{\rm d}x\fr{1}{[x(1-x)p^2+q_0^2]^{1/2+\e}}
 \;, \label{pizero}\\
 \bar{\Pi}_0(p_0=0,p)&=&T\Lambda^{2\e}
 \fr{2^{1+2\e}\sqrt{\pi}}{(4\pi)^{3/2-\e}}
 \fr{\Gamma(3/2+\e)\Gamma(-1/2-\e)}{\Gamma(-\e)}\fr{1}{(p^2)^{3/2+\e}}
 \;, \\
 \widetilde{\Pi}(p_0=0,p)&=&\fr{\beta T \Lambda^{2\e}}{(4\pi)^{3/2-\e}}
 \fr{\Gamma(-1/2+2\e)}{\Gamma(\e)}
 \sum_{q_0}\int_0^1\!\!{\rm d}x
 \fr{(1-x)^{-1+\e}}{[x(1-x)p^2+q_0^2]^{-1/2+2\e}}
 \;, \hspace*{0.7cm}
\ea
which we then substitute into the definitions 
of the sum-integrals in question.

For $S_1^\rmi{II,b}$, we obtain in the fashion described above
\ba
S_1^\rmi{II,b}&=&T^3\Lambda^{6\e}\fr{2^{1+2\e}\sqrt{\pi}}{(4\pi)^{3-2\e}}
  \fr{\Gamma(3/2+\e)\Gamma(1/2+\e)\Gamma(-1/2-\e)}{\Gamma(-\e)}
 \times \nn &\times&
 \sum_{q_0}\int_0^1\!\!{\rm d}x \int\! 
 \fr{{\rm d}^{3-2\e} \vec{p}}{(2\pi)^{3-2\e}}
  \fr{1}{(p^2)^{3/2+\e}[x(1-x)p^2+q_0^2]^{1/2+\e}}
 \nn &=&
 T^3\Lambda^{6\e}\fr{2^{1+2\e}\sqrt{\pi}}{(4\pi)^{3-2\e}}
 \fr{\Gamma(3/2+\e)\Gamma(1/2+\e)\Gamma(-1/2-\e)}
 {\Gamma(-\e)}\bigg\{\sum_{q_0} (q_0^2)^{-1/2-3\e}\bigg\}
 \times \nn &\times&
 \bigg\{\int_0^1\!\!{\rm d}x[x(1-x)]^{2\e}\bigg\}
 \bigg\{\fr{2\pi^{3/2-\e}}{\Gamma(3/2-\e)}
  \fr{1}{(2\pi)^{3-2\e}}\int_0^{\infty}\!\! 
 {\rm d}p \fr{p^{-1-4\e}}{(p^2+1)^{1/2+\e}}\bigg\}
 \;,
\ea
where in the last form we have through rescalings of integration 
variables reduced the sum-integral into a product of sums and integrals 
that can each be trivially evaluated analytically. Performing this task, 
the final result for the function reads
\ba
 S_1^\rmi{II,b}&=&-\fr{T^2}{6(4\pi)^4}
 \biggl(\fr{1}{\e}+3\,\ln\fr{\Lambda^2}{4\pi T^2}-2+3\gammaE\biggr)
 +\rmO(\e) \;.
\ea

To evaluate $S_1^\rmi{III,b}$, 
we note that a calculation exactly parallel to the above produces
\ba
S_1^\rmi{III,b}&=&2{\cal I}_1 T^2\Lambda^{4\e}
 \fr{\Gamma(1/2+\e)}{(4\pi)^{3/2-\e}}
 \sum_{q_0}\int_0^1\!\!{\rm d}x\int\! 
 \fr{{\rm d}^{3-2\e} \vec{p}}{(2\pi)^{3-2\e}}
 \fr{1}{p^4[x(1-x)p^2+q_0^2]^{1/2+\e}}
 \nn &=&
 2{\cal I}_1 T^2\Lambda^{4\e}
 \fr{\Gamma(1/2+\e)}{(4\pi)^{3/2-\e}}\bigg\{
 \sum_{q_0} (q_0^2)^{-1-2\e}\bigg\}
 \bigg\{\int_0^1\!\!{\rm d}x[x(1-x)]^{1/2+\e}\bigg\}
 \times \nn & \times &
 \bigg\{\fr{2\pi^{3/2-\e}}{\Gamma(3/2-\e)}
  \fr{1}{(2\pi)^{3-2\e}}\int_0^{\infty}\!\! 
  {\rm d}p \fr{p^{-2-2\e}}{(p^2+1)^{1/2+\e}}\bigg\} 
 \nn &=&
 -\fr{T^2}{(4\pi)^4}\fr{\pi^2}{36}
 + \rmO(\e)
 \;.
\ea

For $S_1^\rmi{IV,b}$, we finally get
\ba
 S_1^\rmi{IV,b}&=&
 \fr{\beta T^2 \Lambda^{4\e}}{(4\pi)^{3/2-\e}}
 \fr{\Gamma(-1/2+2\e)}{\Gamma(\e)}
 \sum_{q_0}\int_0^1\!\!{\rm d}x
 \int\! \fr{{\rm d}^{3-2\e} \vec{p}}{(2\pi)^{3-2\e}}
 \fr{(1-x)^{-1+\e}}{p^4[x(1-x)p^2+q_0^2]^{-1/2+2\e}}
 \nn &=&
 \fr{\beta T^2 \Lambda^{4\e}}{(4\pi)^{3/2-\e}}
 \fr{\Gamma(-1/2+2\e)}{\Gamma(\e)}\bigg\{\sum_{q_0} (q_0^2)^{-3\e}\bigg\}
 \bigg\{\int_0^1\!\!{\rm d}x\, x^{1/2+\e}(1-x)^{-1/2+2\e}\bigg\}
 \times \nn &\times&
 \bigg\{\fr{2\pi^{3/2-\e}}{\Gamma(3/2-\e)}\fr{1}{(2\pi)^{3-2\e}}
 \int_0^{\infty}\!\! {\rm d}p \fr{p^{-2-2\e}}{(p^2+1)^{-1/2+2\e}}\bigg\}
 \nn &=&
 \fr{T^2}{6(4\pi)^4}\biggl[ \fr{1}{\e}+3\,\ln\fr{\Lambda^2}{4\pi T^2}
 +1-3\gammaE+6\,\ln\(2\pi\)\biggr]
 + \rmO(\e)
 \;.
\ea

\subsubsection{$S_1^\rmi{III,c}$, $S_1^\rmi{IV,c}$ and $S_1^\rmi{IV,d}$}

The last three sum-integrals are trivial to evaluate, as their 
analytic values can be obtained by a straightforward application 
of Eq.~(\ref{in}). The sum of the three gives
\ba
 S_1^\rmi{III,c} + S_1^\rmi{IV,c} + S_1^\rmi{IV,d} 
 &=& 
 3\beta {\cal I}_1{\cal I}_{2+\e} + \widetilde{\beta}{\cal I}_{1+2\e}
 \nn &=&
 \fr{T^2}{8(4\pi)^4}\Bigg\{\fr{1}{\e^2}
 +\fr{1}{\e}\bigg[3\,\ln\fr{\Lambda^2}{4\pi T^2}
 +\fr{17}{6}+\gammaE+2\fr{\zeta'(-1)}{\zeta(-1)}\bigg] 
 + \nn &+&
 \fr{9}{2}\(\ln\fr{\Lambda^2}{4\pi T^2}\)^2
 +\(\fr{17}{2}+3\gammaE+6\fr{\zeta '(-1)}{\zeta(-1)}\)
 \ln\fr{\Lambda^2}{4\pi T^2}
 +\fr{83}{12}
 + \nn &+&
 \fr{13\pi^2}{12} +
 \fr{7\gammaE}{2}-\fr{15\gammaE^2}{2}
 +(5+2\gammaE)\fr{\zeta '(-1)}{\zeta(-1)}
 +2\fr{\zeta ''(-1)}{\zeta(-1)}-16\,\gamma_1\Bigg\}
 \nn & + & 
 \rmO(\e) \;,
\ea
where $\gamma_1$ refers to the first Stieltjes gamma constant
(cf.\ the explanation following \eq\nr{bE3}).

\subsubsection{The full result for $S_1$}

Collecting all the parts above, 
the final result for the sum-integral $S_1$ is seen to read
\ba
 S_1&=&\fr{T^2}{8(4\pi)^4}
 \Bigg\{\fr{1}{\e^2}+\fr{1}{\e}\bigg[3\,\ln\fr{\Lambda^2}{4\pi T^2}
 +\fr{17}{6}+\gammaE+2\fr{\zeta'(-1)}{\zeta(-1)}\bigg]
 + \nn &+&
 \fr{9}{2}\(\ln\fr{\Lambda^2}{4\pi T^2}\)^2+\(\fr{17}{2}
 +3\gammaE+6\fr{\zeta '(-1)}{\zeta(-1)}\)\ln\fr{\Lambda^2}{4\pi T^2}
 +\fr{131}{12}+\fr{31\pi^2}{36} + 8\,\ln (2\pi)
 - \nn & - &
 \fr{9\gammaE}{2}-\fr{15\gammaE^2}{2}
 +(5+2\gammaE)\fr{\zeta '(-1)}{\zeta(-1)}
 +2\fr{\zeta ''(-1)}{\zeta(-1)}-16\,\gamma_1
 +0.388594531408(4) +\rmO(\e) \Bigg\}
 \;. \nn \la{S1}
\ea
Evaluating the non-logarithmic terms numerically leads 
to the result in \eq\nr{S1res}.

%%%%%%%%%%%%%%%%%%%%%%%%%%%% SUBSECTION %%%%%%%%%%%%%%%%%%%%%%%%%%%%%%%%%%
%
\subsection{Strategy for determining $S_2$}

As with $S_1$, we begin the evaluation of our 
only genuinely 4-loop sum-integral $S_2$, 
defined in \eq\nr{S2def}, by dividing 
it into several pieces that are then calculated separately.
Defining a regularized, or subtracted, zero-temperature limit 
for the function $\Pi(P)$ by
\ba
 \Pi^{(0)}_s(P)&\equiv& \Pi^{(0)}(P)-\fr{1}{(4\pi)^2\e}
 \nn &=&
 -\fr{1}{(4\pi)^2}\bigg\{\ln\fr{P^2}{(2\pi T)^2}
 -\ln\fr{\Lambda^2}{4\pi T^2}-2-2\,\ln\,2+\gammaE\bigg\}
 +{\mathcal O}(\e)
 \nn &\equiv&
 -\fr{1}{(4\pi)^2}\bigg\{\ln\fr{P^2}{(2\pi T)^2}+\kappa\bigg\}
 +{\mathcal O}(\e)\;, \la{defPis}
\ea
we write
$
 \Pi(P) = 1/(4\pi)^2\e + \Pi^{(0)}_s(P) + \Pi^{(T)}(P)
$, 
and drop terms that vanish in dimensional regularization. This leads to
\ba
 S_2&=&\sumint_P \[\Pi^{(T)}(P)\]^3
   +3\,\sumint_P \[\Pi^{(T)}(P)\]^2\Pi^{(0)}_s(P)
   +3\,\sumint_P \Pi^{(T)}(P)\[\Pi^{(0)}_s(P)\]^2
  + \nn &+&
   \sumint_P \[\Pi^{(0)}_s(P)\]^3
  -\fr{3}{(4\pi)^4\e^2}\sumint_P\Pi(P)
 \nn &\equiv& 
 S_2^\rmi{I} + S_2^\rmi{II} + S_2^\rmi{III} + S_2^\rmi{IV} + S_2^\rmi{V}
 \;.
\ea
Here, we again decompose the various 
sum-integrals into further pieces according to
\ba
 S_2^\rmi{I} &=& \sumint '_P \[\Pi^{(T)}(P)\]^3
 + T\Lambda^{2\e}\!\!\int\! \fr{{\rm d}^{3-2\e} \vec{p}}{(2\pi)^{3-2\e}}
 \[\Pi^{(T)}(p_0=0,p)\]^3
 \nn &\equiv& S_2^\rmi{I,a} + S_2^\rmi{I,b}
 \;,\\
 S_2^\rmi{II} &=& 3\,\sumint_P 
 \bigg\{\[\Pi^{(T)}(P)\]^2-\fr{4\({\cal I}_1\)^2}{P^4}\bigg\}\Pi^{(0)}_s(P)
 +12\({\cal I}_1\)^2\sumint_P \fr{\Pi^{(0)}_s(P)}{P^4}\nn
 &\equiv& S_2^\rmi{II,a} + S_2^\rmi{II,b}
 \;,\\
 S_2^\rmi{III} &=& 3\,\sumint_P \Delta'\Pi(P)\[\Pi^{(0)}_s(P)\]^2
 +6{\cal I}_1\sumint_P\fr{\[\Pi^{(0)}_s(P)\]^2}{P^2}
 + \nn &+&
 \fr{24T^4J_1}{3-2\e}\sumint '_P\[\Pi^{(0)}_s(P)\]^2
 \biggl[ \fr{1}{P^4}-(4-2\e)\fr{p_0^2}{P^6}\biggr]
 \nn &\equiv& 
 S_2^\rmi{III,a} + S_2^\rmi{III,b}+S_2^\rmi{III,c}
 \;.
\ea
% in which we have suppressed a factor $\(1-\delta_{p_0,0}\)$
% in the definition of $S_2^{III,c}$, as it --- due to the vanishing
% of scaleless integrals in dimensional regularization
% --- in any case would not affect the result.
This time we classify the various parts as
\begin{itemize}
\item Finite (but complicated) terms that can be evaluated numerically: 
 $S_2^\rmi{I,a}$, $S_2^\rmi{II,a}$, $S_2^\rmi{III,a}$.
\item Possibly divergent terms, whose divergent parts can be evaluated 
 analytically and finite parts numerically: $S_2^\rmi{I,b}$.
\item Terms that are trivial to compute due to Eq.~(\ref{in}): 
 $S_2^\rmi{II,b}$, $S_2^\rmi{III,b}$, $S_2^\rmi{III,c}$, 
 $S_2^\rmi{IV}$, $S_2^\rmi{V}$.
\end{itemize}
We move on to present the evaluation of the first four sum-integrals in detail.

\subsubsection{$S_2^\rmi{I,a}$}

The term $S_2^\rmi{I,a}$ is entirely finite in three dimensions, 
so we set $\e=0$ and proceed to its numerical evaluation. There 
are two comparably simple ways of doing this, as one can either 
work entirely in coordinate space, or use the relation
\ba
 \Pi^{(T)}(P) &=& \fr{T}{4\pi p}\int_0^{\infty}\!\!{\rm d}r\, 
  \fr{\sin pr}{r}\(\coth \bar{r}-\fr{1}{\bar{r}}\)e^{-|p_0|r}
 \;, \la{repPiT}
\ea
and perform the sum-integration over $P$ directly in momentum space. 
Here, we choose the latter approach, but have verified the result 
also by the former method.
Scaling all variables dimensionless, we straightforwardly get
\ba
 S_2^\rmi{I,a}&=&\fr{T^4}{(4\pi)^4}\fr{4}{\pi}
 \sum_{n=1}^{\infty}\int_0^{\infty}\!\!{\rm d}p \,
 \fr{1}{p}\bigg\{\int_0^{\infty}\!\!{\rm d}r \,\fr{\sin(pr)}{r}
 \(\coth r-\fr{1}{r}\)e^{-nr}\bigg\}^3
 \nn &\approx& 
 \fr{T^4}{(4\pi)^4}\times 0.0092313322549(1)
 \;. \la{resS2Ia}
\ea
The summation can be accelerated by noting that for large $n$, 
the integral inside the curly brackets yields $p/3(p^2 + n^2)$, 
and the subsequent $p$-integral $\pi/432 n^3$. This leading
behaviour can be subtracted and the corresponding sum
carried out analytically. 

\subsubsection{$S_2^\rmi{II,a}$}

The only potential divergence in $S_2^\rmi{II,a}$ is of an IR type, 
and produced by the zero-mode piece $p_0 = 0$ of the $1/P^4$ subtraction term.
However, there is in fact no divergence in dimensional regularization, 
given that the zero-mode piece does not contribute: 
$
 \int \! {\rm d}^{3-2\e}\vec{p} \, [\Pi^{(0)}(p_0=0,p) -1/(4\pi)^2\e]/p^4 = 0 
$.
We may therefore set $\e=0$ from the beginning, 
if we agree to throw away the subtraction term for $p_0=0$. 
Using furthermore the result of 
\eqs(2.32) and (2.34) of Ref.~\cite{az},
\ba
 \sumint_P \bigg\{\[\Pi^{(T)}(P)\]^2-\fr{4\({\cal I}_1\)^2}{P^4}\bigg\}
 =\fr{1}{(4\pi)^2}\(\fr{T^2}{12}\)^{\!\!2}
 \biggl[\fr{28}{15}-8\gammaE
 +24\fr{\zeta'(-1)}{\zeta(-1)}-16\fr{\zeta'(-3)}{\zeta(-3)}\biggr]
 \;, \nn
\ea
as well as the representations for $\Pi^{(0)}_s$ and $\Pi^{(T)}$
as in \eqs\nr{defPis}, \nr{repPiT}, we thus obtain
\ba
 S_2^\rmi{II,a}&=&-\fr{T^4}{(4\pi)^4}\fr{3}{\pi}
  \sum_n \int_0^{\infty}\!\!{\rm d}p\, \ln(p^2+n^2)
  \times \nn &\times&
 \bigg\{\[\int_0^{\infty}\!\!{\rm d}r \,\fr{\sin(pr)}{r}
 \(\coth r-\fr{1}{r}\)e^{-|n|r}\]^{ 2}
 -\fr{p^2}{9}\fr{1-\delta_{n}}{(p^2+n^2)^2}\bigg\}
 - \nn &-&
 \fr{3\kappa}{(4\pi)^4}\(\fr{T^2}{12}\)^{\!\!2}
  \biggl[\fr{28}{15}-8\gammaE
  +24\fr{\zeta'(-1)}{\zeta(-1)}-16\fr{\zeta'(-3)}{\zeta(-3)}\biggr]
 \;. 
\ea
For large $n$, the expression inside the curly brackets approaches
$
 4 p^2 (p^2 - 3 n^2)/135 (p^2+n^2)^4
$, 
and the subsequent $p$-integral yields
$
 \pi [7 - 12 \ln(2 n)]/ 3240 n^3
$. 
This term can again be subtracted and the corresponding
sums carried out analytically, 
to accelerate the convergence of the remaining numerical 
summation. We thus get 
\ba
 S_2^\rmi{II,a}&=&
 \fr{T^4}{(4\pi)^4}\Bigg\{\fr{1}{48}
 \(\ln\fr{\Lambda^2}{4\pi T^2}+2+2\,\ln\,2-\gammaE\)
 \[\fr{28}{15}-8\gammaE
 +24\fr{\zeta'(-1)}{\zeta(-1)}-16\fr{\zeta'(-3)}{\zeta(-3)}\]
 \nn
 &+&2.0344721052(1)\Bigg\}
 \nn &=&\fr{T^4}{(4\pi)^4}
 \Bigg\{\[\fr{7}{180}-\fr{\gammaE}{6}
 +\fr{1}{2}\fr{\zeta'(-1)}{\zeta(-1)}
 -\fr{1}{3}\fr{\zeta'(-3)}{\zeta(-3)}\]\ln\fr{\Lambda^2}{4\pi T^2}
 +4.0572056435(1)\Bigg\}
 \;. \nn 
\ea

\subsubsection{$S_2^\rmi{III,a}$}

With $S_2^\rmi{III,a}$, we begin the calculation by noting that if we write
\ba
\(\Pi^{(0)}_s(P)\)^2&=&\fr{1}{(4\pi)^4}
 \bigg\{\(\ln\fr{P^2}{(2\pi T)^2}\)^{\!\!2}-\kappa^2\bigg\}
 -\fr{2\kappa}{(4\pi)^2}\Pi^{(0)}_s(P)
 \;,
\ea
and use the result in \eq(D.14) of Ref.~\cite{az}, 
\ba
 \sumint_P \Delta'\Pi(P)\Pi^{(0)}_s(P)
 &=& -\fr{1}{(4\pi)^2}\(\fr{T^2}{12}\)^{\!\!2}
 \bigg\{\fr{46}{15}-\fr{12}{5}\gammaE+4\fr{\zeta'(-1)}{\zeta(-1)}
 -\fr{8}{5}\fr{\zeta'(-3)}{\zeta(-3)}\bigg\}
 \;,  \hspace*{0.5cm}
\ea
as well as that in Eq.~(\ref{in})
of our Sec.~\ref{se:betaE1}, in order to show that 
$
 {\Sigma_P}\!\!\!\!\!\!\!\!\raise0.3ex\hbox{$\int$}\;\; \Delta' \Pi(P) = 
 {\Sigma_P}\!\!\!\!\!\!\!\!\raise0.3ex\hbox{$\int$}\;\;
 [\Pi(P) - \Pi^{(0)}(P) - ...] = \rmO(\e) 
$, 
so that the $\kappa^2$-term can be neglected, 
we can immediately reduce the evaluation of the original 
sum-integral into that of the simpler function
\ba
\sumint_P \Delta'\Pi(P)\(\ln\fr{P^2}{(2\pi T)^2}\)^{\!\!2}.
\ea
In performing the latter task, we again have a choice between 
working entirely in coordinate space or combining 
(spatial) coordinate and momentum space techniques;
this time we for illustration choose the former approach.

We begin the calculation by recalling the result 
of Eq.~(\ref{epsint}), from which we obtain by differentiation 
with respect to $\alpha$
\ba
 \int\fr{{\rm d}^3\vec{p}}{(2\pi)^3}
 e^{-i\mathbf{p}\cdot\mathbf{r}}\[\ln(p^2+m^2)\]^2&=&
 -\fr{e^{-|m|r}}{\pi r^3}
  \bigg\{2-(1+|m|r)\bigg(\gammaE-\ln\fr{2|m|}{r}\bigg)\bigg\}
 - \nn &-&
 \fr{e^{|m|r}}{\pi r^3}(1-|m|r){\rm E_1}(2|m|r)
 \nn &\equiv&
 \fr{e^{-|m|r}}{\pi r^3}f(r,m)
 \;.
\ea
Here, we have used the relation
\ba
 \bigg[\fr{\partial}{\partial \nu} {\rm K}_\nu (z)\bigg]_{\nu=3/2}
 &=&\sqrt{\fr{\pi}{2z^3}}e^{-z}\Big\{2-(z-1)\,e^{2z}\,{\rm E_1}(2z)\Big\}
 \;,
\ea
in which the exponential integral E$_1$ 
is related to the exponential integral function Ei through 
${\rm E}_1(z)=-{\rm Ei}(-z)$. 
We also note that for $m=0$, the function $f(r,m)$ becomes
\ba
 f(r,0)&=&2\,( \ln\,r- 1 + \gammaE) 
 \;.
\ea

Going now to coordinate space in the sum-integral 
and scaling the integration variables dimensionless, we obtain
\ba
 \sumint_P \Delta'\Pi(P)\(\ln\fr{P^2}{(2\pi T)^2}\)^{\!\!2}
 &=&T^4\sum_n \int_0^{\infty}\!\!{\rm d}r
 \,\fr{1}{r^3}f(r,n)
 \times \nn & & \times 
 \(\coth r-\fr{1}{r}-\fr{r}{3}
 +\(1-\delta_{n}\)\fr{r^3}{45}\)e^{-2|n|r}
 \nn &=&
  2T^4\bigg\{\int_0^{\infty}\!\!{\rm d}r\,\fr{1}{r^3}
  \(\ln\,r-1+\gammaE\)\(\coth\,r-\fr{1}{r}-\fr{r}{3}\)
 + \nn &+&
 \sum_{n=1}^\infty \int_0^{\infty}\!\!{\rm d}r
 \,\fr{1}{r^3}f(r,n)
 \(\coth\,r-\fr{1}{r}-\fr{r}{3}+\fr{r^3}{45}\)e^{-2nr}
 \bigg\}
 \;, \nn
\ea
where all terms are both IR and UV convergent thanks to 
the $\(1-\delta_{n}\)$ factor above. 
Given that it is perfectly finite,
the zero-mode integral 
is furthermore simple to evaluate through a procedure introduced 
in Ref.~\cite{az}:
one temporarily introduces a convergence factor $r^{\alpha}$ 
into the integrand, writes $\coth r = 1+2/(e^{2r}-1)$,
throws out all terms that are simple powers of $r$, performs 
the remaining integrals and finally proceeds to 
the limit $\alpha\rightarrow 0$ (which result can also be reproduced
numerically). 
A straightforward calculation employing
numerical integration for the non-zero mode terms then yields
\ba
\sumint_P \Delta'\Pi(P)
 \(\ln\fr{P^2}{(2\pi T)^2}\)^{\!\!2}&=&
 2T^4\bigg\{\fr{1}{\pi^2}\[(1-\gammaE-\ln\,\pi)\zeta(3)+\zeta'(3)\]
 + \nn &+&
 \sum_{n=1}^\infty \int_0^{\infty}\!\!{\rm d}r
 \,\fr{1}{r^3}f(r,n)\(\coth\,r-\fr{1}{r}-\fr{r}{3}+\fr{r^3}{45}\)
 e^{-2nr}\bigg\}
 \nn &=&
 T^4\bigg\{\fr{2}{\pi^2}\[(1-\gammaE-\ln\,\pi)\zeta(3)+\zeta'(3)\]
 -0.002580375031(1)\bigg\}\nn
 &\approx&-T^4\times 0.218586170715(1)
 \;.
\ea
Like before, the large-$n$ behaviour can be worked out analytically
and subtracted, to accelerate the convergence of the summation: it is 
$
 2 T^4 [11 - 24 \ln(2 n)]/ 9072 n^3
$.

Collecting everything together, we finally have
\ba
 S_2^\rmi{III,a}&=&
 -\fr{T^4}{(4\pi)^4}\Bigg\{\fr{1}{24}
 \(\ln\fr{\Lambda^2}{4\pi T^2}+2+2\,\ln\,2-\gammaE\)
 \times \nn & \times & 
 \[\fr{46}{15}-\fr{12}{5}\gammaE+4\fr{\zeta'(-1)}{\zeta(-1)}
 -\fr{8}{5}\fr{\zeta'(-3)}{\zeta(-3)}\]
 - \nn &-&
 \fr{6}{\pi^2}\[(1-\gammaE-\ln\,\pi)\zeta(3)+\zeta'(3)\]
 +0.007741125093(3)\Bigg\}\nn
 &=&-\fr{T^4}{(4\pi)^4}
 \Bigg\{\[\fr{23}{180}-\fr{\gammaE}{10}
 +\fr{1}{6}\fr{\zeta'(-1)}{\zeta(-1)}
 -\fr{1}{15}\fr{\zeta'(-3)}{\zeta(-3)}\]\ln\fr{\Lambda^2}{4\pi T^2}
 + \nn & + & 1.661043190056(3) \Bigg\}
 \;. 
\ea

\subsubsection{$S_2^\rmi{I,b}$}

The function $S_2^\rmi{I,b}$ has one unique property: 
while entirely UV finite, it is the only sum-integral 
encountered in the present calculation that has a non-trivial
logarithmic IR divergence that
does not vanish in dimensional regularization. 
To see this, we note that the contribution of 
the Matsubara zero-mode to $\Pi^{(T)}(p_0=0,p)$ 
[the superscript $(T)$ is irrelevant here] reads
\ba
\Pi_{\rm IR}^{(T)}(p_0=0,p)&\equiv&
 T\Lambda^{2\e}\!\!\int\! \fr{{\rm d}^{3-2\e} \vec{q}}{(2\pi)^{3-2\e}}
 \fr{1}{q^2(\mathbf{q}-\mathbf{p})^2}
 %\nn &=&
 %\fr{T\Lambda^{2\e}}{(4\pi)^{3/2-\e}}\Gamma(1/2+\e)
 %\int_0^{\infty}\!\!{\rm d}x (x(1-x))^{-1/2-\e}\fr{1}{(p^2)^{1/2+\e}}
 \nn &=&
 T\Lambda^{2\e}\fr{2^{2\e}\sqrt{\pi}}{(4\pi)^{3/2-\e}}
 \fr{\Gamma(1/2+\e)\Gamma(1/2-\e)}{\Gamma(1-\e)}\fr{1}{(p^2)^{1/2+\e}}
 \nn &\equiv& 
 T\Lambda^{2\e}\fr{A_{\e}}{8(p^2)^{1/2+\e}}
 \;\;=\;\;\fr{T}{8p}+{\mathcal O}(\e) \;,
\ea
which leads to an IR singularity in the integral defining 
$S_2^\rmi{I,b}$ but cannot be separated from it 
without causing new UV divergences.

To facilitate an analytic evaluation of the divergence, 
we add and subtract from the integrand of $S_2^\rmi{I,b}$ a term of the form
$\[\Pi_{\rm IR}^{(T)}(p_0=0,p)\]^3\fr{m^2}{p^2+m^2}$, where the 
in principle arbitrary regularization mass $m$ 
is chosen to be $m=2\pi T$ for computational convenience.
This enables us to write
\ba
 S_2^\rmi{I,b}&=&
 T\!\int\! \fr{{\rm d}^{3} \vec{p}}{(2\pi)^{3}}
 \bigg\{\[\Pi^{(T)}(p_0=0,p)\]^{\!3}-
 \(\fr{T}{8p}\)^{\!\!3}\fr{(2\pi T)^2}{p^2+(2\pi T)^2}\bigg\}
 + \nn &+&
 T\Lambda^{2\e}\!\!\int\! \fr{{\rm d}^{3-2\e} \vec{p}}{(2\pi)^{3-2\e}}
 \[\Pi_{\rm IR}^{(T)}(p_0=0,p)\]^{\!3}\fr{(2\pi T)^2}{p^2+(2\pi T)^2}
 + \rmO(\epsilon)
 \;,
\ea
where the first piece has the virtue of being finite at $d=3$, 
while the second one is straightforward to evaluate analytically. 
Substituting here the Fourier representation of $\Pi^{(T)}(P)$ 
as well as the above form of $\Pi_{\rm IR}^{(T)}(p_0=0,p)$, 
we obtain after scaling all variables dimensionless
\ba
 S_2^\rmi{I,b}&=&
 \fr{T^4}{64(4\pi)^2}\!\int_0^{\infty}\!\!{\rm d}p\,\fr{1}{p}
 \bigg\{\[\fr{2}{\pi}\int_0^{\infty}\!\!{\rm d}r\, \fr{\sin(pr)}{r}
 \(\coth r-\fr{1}{r}\)\]^{3}-\fr{1}{p^2+1}\bigg\}
 + \nn &+&
 \fr{T^4\Lambda^{8\e}A_{\e}^3}{512}\fr{(2\pi T)^{-8\e}}{(2\pi)^{3-2\e}}
 \fr{2\pi^{3/2-\e}}{\Gamma(3/2-\e)}
 \int_0^{\infty}\!\!{\rm d}p\,\fr{p^{-1-8\e}}{p^2+1}
 + \rmO(\epsilon)
 \nn
 &=&-\fr{T^4}{512(4\pi)^2}
 \Bigg\{\fr{1}{\e}+4\,\ln\fr{\Lambda^2}{4\pi T^2}
 +2+12\,\ln\,2-4\gammaE+20.91614600219(1) + \rmO(\e)\Bigg\}\nn
 &=&-\fr{T^4}{512(4\pi)^2}\Bigg\{\fr{1}{\e}
 +4\,\ln\fr{\Lambda^2}{4\pi T^2}
 +28.92504950930(1)+\rmO(\e)\Bigg\}
 \;. \la{S2Ib}
\ea

\subsubsection{$S_2^\rmi{II,b}$, $S_2^\rmi{III,b}$, 
$S_2^\rmi{III,c}$, $S_2^\rmi{IV}$ and $S_2^\rmi{V}$}

The remaining five sum-integrals are again obtained 
through a straightforward application of Eq.~(\ref{in}),
as they can be written in the forms
\ba
 S_2^\rmi{II,b}&=&
 12\({\cal I}_1\)^2\bigg\{\beta{\cal I}_{2+\e}
 -\fr{1}{(4\pi)^2\e}{\cal I}_2\bigg\}
 \;,\\
 S_2^\rmi{III,b}&=&
 6{\cal I}_1\bigg\{\beta^2{\cal I}_{1+2\e}
 -\fr{2\beta}{(4\pi)^2\e}{\cal I}_{1
 +\e}+\fr{1}{(4\pi)^4\e^2}{\cal I}_{1}\bigg\}
 \;,\\
 S_2^\rmi{III,c}&=&
 \fr{24J_1T^4}{3-2\e}\bigg\{\beta^2{\cal I}_{2+2\e}
 -\fr{2\beta}{(4\pi)^2\e}{\cal I}_{2+\e}+\fr{1}{(4\pi)^4\e^2}{\cal I}_{2}
 - \nn &-&(
 4-2\e)\[ \beta^2{\cal I}^2_{3+2\e}-
 \fr{2\beta}{(4\pi)^2\e}{\cal I}^2_{3+\e}
 +\fr{1}{(4\pi)^4\e^2}{\cal I}^2_{3} \]\bigg\}
 \;, \\
 S_2^\rmi{IV}&=& 
 \beta^3{\cal I}_{3\e}
 -\fr{3\beta^2}{(4\pi)^2\e}{\cal I}_{2\e}
 +\fr{3\beta}{(4\pi)^4\e^2}{\cal I}_\e
 \;,\\
 S_2^\rmi{V} &=&-\fr{3}{(4\pi)^4\e^2}\({\cal I}_1\)^2 
 \;,
\ea
the sum of which gives
\ba
 &&\!\!\!\!\!\!\!\!\!\!
 S_2^\rmi{II,b} + S_2^\rmi{III,b} + S_2^\rmi{III,c} 
 + S_2^\rmi{IV} + S_2^\rmi{V}\nn
 &=&-\fr{T^4}{16(4\pi)^4}\Bigg\{\fr{1}{\e^2}+\fr{1}{\e}
 \bigg[2\,\ln\fr{\Lambda^2}{4\pi T^2}+\fr{10}{3}-2\gammaE
 +4\fr{\zeta'(-1)}{\zeta(-1)}\bigg]
 +\(\ln\fr{\Lambda^2}{4\pi T^2}\)^{\!\!2}
 - \nn &-&
 \[\fr{2}{9}+\fr{46\gammaE}{15}-\fr{16}{3}\fr{\zeta '(-1)}{\zeta(-1)}
  +\fr{4}{15}\fr{\zeta '(-3)}{\zeta(-3)}\]\ln\fr{\Lambda^2}{4\pi T^2}
 -\fr{193}{27}+\fr{\pi^2}{6}
 -\fr{74\gammaE}{45}
 + \nn &+&
 \fr{31\gammaE^2}{15}  + 
 \fr{8(1-2\gammaE)}{3}\fr{\zeta '(-1)}{\zeta(-1)}
 +4\(\fr{\zeta'(-1)}{\zeta(-1)}\)^{\!\!2}
 -\fr{4(14-3\gammaE)}{45}\fr{\zeta '(-3)}{\zeta(-3)}
 + \nn & + & 
  \fr{4}{3}\fr{\zeta ''(-1)}{\zeta(-1)} 
 -\fr{4}{15}\fr{\zeta''(-3)}{\zeta(-3)}
 +\fr{32}{15}\,\gamma_1\Bigg\}
 + \rmO(\e)
 \;. 
\ea

\subsubsection{The full result for $S_2$}

Taking the sum of all the various pieces of $S_2$ displayed above, 
we can now finally write down the result for 
the entire sum-integral in the form
\ba
 S_2&=&
 -\fr{T^4}{16(4\pi)^4}\Bigg\{\fr{1}{\e^2}
 +\fr{1}{\e}\bigg[2\,\ln\fr{\Lambda^2}{4\pi T^2}+\fr{10}{3}
 -2\gammaE+4\fr{\zeta'(-1)}{\zeta(-1)}\bigg]
 +\(\ln\fr{\Lambda^2}{4\pi T^2}\)^{\!\!2}
 + \nn &+&
 \[\fr{6}{5}-2\gammaE+4\fr{\zeta '(-3)}{\zeta(-3)}\]
 \ln\fr{\Lambda^2}{4\pi T^2}
 -\fr{581}{135}+\fr{\pi^2}{6}+\fr{32(4+3\gammaE)}{45}\ln 2
 -\fr{14\gammaE}{15}+\gammaE^2
 - \nn &-&
 \fr{8(3+4\,\ln\,2)}{3}\fr{\zeta '(-1)}{\zeta(-1)}
 +4\(\fr{\zeta'(-1)}{\zeta(-1)}\)^{\!\!2}
 +\(\fr{328}{45}+\fr{128\,\ln\,2}{15}-4\gammaE\)\fr{\zeta '(-3)}{\zeta(-3)}
 - \nn &-&
 \fr{96}{\pi^2}\[(1-\gammaE-\ln\,\pi)\zeta(3)+\zeta'(3)\]
 +\fr{4}{3}\fr{\zeta ''(-1)}{\zeta(-1)}
 -\fr{4}{15}\fr{\zeta''(-3)}{\zeta(-3)}
 + \nn & + & 
 \fr{32}{15}\,\gamma_1
 -32.575396998(2)\Bigg\}
 - \nn &-&
 \fr{T^4}{512(4\pi)^2}\Bigg\{\fr{1}{\e}
 +4\,\ln\fr{\Lambda^2}{4\pi T^2}+2
 +12\,\ln\,2-4\gammaE
 +20.91614600219(1)\Bigg\}
 + \rmO(\e) \;. \nn 
 \la{S2}
\ea
On the last row, we have separated the IR-singular contribution from
$S_2^\rmi{I,b}$ in \eq\nr{S2Ib}.
Evaluating the non-logarithmic terms numerically leads to the 
result in \eq\nr{S2res}.

% \newpage

%%%%%%%%%%%%%%%%%%%%%%%%% BIBLIO %% REFERENCES %%%%%%%%%%%%%%%%%%%%%%%%%%%%%%


\begin{thebibliography}{99}


\bibitem{az} 
  P.~Arnold and C.~Zhai,
  %{\it The three loop free energy for pure gauge QCD,}
  Phys.\ Rev.\  {D 50} (1994) 7603
  [hep-ph/9408276];
  %%CITATION = HEP-PH 9408276;%%
  %
  %{\it The three loop free energy for high temperature QED 
  %and QCD with fermions,}
  {\it ibid.}\  {51} (1995) 1906
  [hep-ph/9410360].
  %%CITATION = HEP-PH 9410360;%%

\bibitem{zk}
  C.~Zhai and B.~Kastening,
  %{\it The free energy of hot gauge theories with fermions through $g^5$,}
  Phys.\ Rev.\  {D 52} (1995) 7232 [hep-ph/9507380].
  %%CITATION = HEP-PH 9507380;%%

\bibitem{bn2}
  E. Braaten and A. Nieto,
  %{\it Free energy of QCD at high temperature,}
  Phys.\ Rev.\ D 53 (1996) 3421 
  [hep-ph/9510408].
  %%CITATION = HEP-PH 9510408;%%

\bibitem{gsixg}
  K.~Kajantie, M.~Laine, K.~Rummukainen and Y.~Schr\"oder,
  %{\it The pressure of hot QCD up to $g^6 \ln (1/g)$},
  Phys.\ Rev.\ D 67 (2003) 105008
  [hep-ph/0211321]. 
  %%CITATION = HEP-PH 0211321;%%

\bibitem{linde}
  A.D.~Linde,
  %{\it Infrared problem in thermodynamics of the Yang-Mills gas,}
  Phys.\ Lett.\ {B 96} (1980) 289.
  %%CITATION = PHLTA,B96,289;%%

\bibitem{gpy}
  D.J.~Gross, R.D.~Pisarski and L.G.~Yaffe,
  %{\it QCD and instantons at finite temperature,}
  Rev.\ Mod.\ Phys.\ {53} (1981) 43.
  %%CITATION = RMPHA,53,43;%%

\bibitem{plaq}
  A.~Hietanen, K.~Kajantie, M.~Laine, K.~Rummukainen and Y.~Schr\"oder,
  %``Plaquette expectation value and gluon condensate in three dimensions,''
  JHEP {01} (2005) 013
  [hep-lat/0412008];
  %%CITATION = HEP-LAT 0412008;%%
%
% \bibitem{hk}
  A.~Hietanen and A.~Kurkela,
  %``Plaquette expectation value and lattice free energy of three-dimensional
  %SU(N(c)) gauge theory,''
  JHEP {11} (2006) 060
  [hep-lat/0609015].
  %%CITATION = JHEPA,0611,060;%%

\bibitem{nspt_mass}
  F.~Di Renzo, M.~Laine, V.~Miccio, Y.~Schr\"oder and C.~Torrero,  
  %``The leading non-perturbative coefficient in the weak-coupling
  %  expansion of hot QCD pressure,''
  JHEP {07} (2006) 026 
  [hep-ph/0605042].
  %%CITATION = HEP-PH 0605042;%%   

\bibitem{Nf}
  G.D.~Moore,
  %``Pressure of hot QCD at large N(f),''
  JHEP {10} (2002) 055
  [hep-ph/0209190];
  %%CITATION = HEP-PH 0209190;%%
%
  A.~Ipp, G.D.~Moore and A.~Rebhan,
  %``Comment on 'Pressure of hot QCD at large N(f)' with corrected exact
  %result,''
  JHEP {01} (2003) 037
  [hep-ph/0301057].
  %%CITATION = HEP-PH 0301057;%%
  
\bibitem{av2}
  A.~Vuorinen,
  %{\it The pressure of QCD at finite temperatures and chemical potentials,}
  Phys.\ Rev.\ D {68} (2003) 054017
  [hep-ph/0305183];
  %%CITATION = HEP-PH 0305183;%%
%
  A.~Ipp, K.~Kajantie, A.~Rebhan and A.~Vuorinen,
  %``The pressure of deconfined QCD for all temperatures and quark chemical
  %potentials,''
  Phys.\ Rev.\  D {74} (2006) 045016
  [hep-ph/0604060].
  %%CITATION = PHRVA,D74,045016;%%

\bibitem{Nfmu}
  A.~Ipp and A.~Rebhan,
  %``Thermodynamics of large-N(f) QCD at finite chemical potential,''
  JHEP {06} (2003) 032
  [hep-ph/0305030];
  %%CITATION = HEP-PH 0305030;%%
%
  A.~Ipp, A.~Rebhan and A.~Vuorinen,
  %``Perturbative QCD at non-zero chemical potential: Comparison with the
  %large-N(f) limit and apparent convergence,''
  Phys.\ Rev.\ D {69} (2004) 077901
  [hep-ph/0311200].
  %%CITATION = HEP-PH 0311200;%%

%\cite{Laine:2006cp}
\bibitem{massdep}
  M.~Laine and Y.~Schr\"oder,
  %``Quark mass thresholds in QCD thermodynamics,''
  Phys.\ Rev.\ D { 73} (2006) 085009
  [hep-ph/0603048].
  %%CITATION = HEP-PH 0603048;%%

\bibitem{gv}
  A.~Gynther and M.~Veps\"al\"ainen,
  %``Pressure of the standard model at high temperatures,''
  JHEP {01} (2006) 060
  [hep-ph/0510375];
  %%CITATION = HEP-PH 0510375;%%
%
  %``Pressure of the standard model near the electroweak phase transition,''
  JHEP {03} (2006) 011
  [hep-ph/0512177].
  %%CITATION = HEP-PH 0512177;%%

\bibitem{aminusb}
  K.~Kajantie, M.~Laine, K.~Rummukainen and Y.~Schr\"oder,
  %``Four-loop vacuum energy density of the SU(N(c)) + adjoint Higgs theory,''
  JHEP {04} (2003) 036
  [hep-ph/0304048].
  %%CITATION = HEP-PH 0304048;%%

\bibitem{bE2}
  Y.~Schr\"oder,
  % {\it Evading the infrared problem of thermal QCD,}
  Proceedings of {\it Strong and Electroweak Matter 2004}, 
  eds.\ K.J.~Eskola {\it et al}, p.~261
  [hep-ph/0410130].
  %%CITATION = HEP-PH 0410130;%%

\bibitem{gE2}
  M.~Laine and Y.~Schr\"oder,
%  {\it Two-loop QCD gauge coupling at high temperatures,}
  JHEP {03} (2005) 067
  [hep-ph/0503061].
  %%CITATION = HEP-PH 0503061;%%

\bibitem{quart}
  S.~Nadkarni,
  %``Dimensional Reduction In Finite Temperature Quantum Chromodynamics. 2,''
  Phys.\ Rev.\ D {38} (1988) 3287;
  %%CITATION = PHRVA,D38,3287;%%
  %
  N.P.~Landsman,
  %``Limitations To Dimensional Reduction At High Temperature,''
  Nucl.\ Phys.\ B {322} (1989) 498.
  %%CITATION = NUPHA,B322,498;%%

%\cite{Frenkel:1992az}
\bibitem{fst}
  J.~Frenkel, A.V.~Saa and J.C.~Taylor,
  %``The Pressure in thermal scalar field theory to three loop order,''
  Phys.\ Rev.\ D {46} (1992) 3670.
  %%CITATION = PHRVA,D46,3670;%%

%\cite{Parwani:1994zz}
\bibitem{ps}
  R.~Parwani and H.~Singh,
  %``The Pressure of hot (g**2 phi**4) theory at order g**5,''
  Phys.\ Rev.\ D { 51} (1995) 4518
  [hep-th/9411065].
  %%CITATION = HEP-TH 9411065;%%

%\cite{Braaten:1995cm}
\bibitem{bn1}
  E.~Braaten and A.~Nieto,
  %``Effective Field Theory Approach To High Temperature Thermodynamics,''
  Phys.\ Rev.\ D { 51} (1995) 6990
  [hep-ph/9501375].
  %%CITATION = HEP-PH 9501375;%%

%\cite{Kastening:1996nj}
\bibitem{bmk}
  B.M.~Kastening,
  %``Four-Loop Vacuum Energy Beta Function in O(N) Symmetric Scalar Theory,''
  Phys.\ Rev.\  D {54} (1996) 3965
  [hep-ph/9604311].
  %%CITATION = PHRVA,D54,3965;%%

\bibitem{dr}
  P. Ginsparg, 
  % {\it First and second order phase transitions 
  % in gauge theories at finite temperature,}
  Nucl.\ Phys.\ B 170 (1980) 388;
  %%CITATION = NUPHA,B170,388;%% 
  %
  T. Appelquist and R.D. Pisarski,
  % {\it High-temperature Yang-Mills theories and three-dimensional 
  % Quantum Chromodynamics,}
  Phys.\ Rev.\ D 23 (1981) 2305.
  %%CITATION = PHRVA,D23,2305;%%

\bibitem{generic}
  K.~Kajantie, M.~Laine, K.~Rummukainen and M.~Shaposhnikov,
  %{\it Generic rules for high temperature dimensional reduction 
  %and their application to the Standard Model,}
  Nucl.\ Phys.\ {B 458} (1996) 90 
  [hep-ph/9508379].
  %%CITATION = NUPHA,B458,90;%%

\bibitem{pert}
  K.~Farakos, K.~Kajantie, K.~Rummukainen and M.E.~Shaposhnikov,
  %``3-D physics and the electroweak phase transition: Perturbation theory,''
  Nucl.\ Phys.\  B {425} (1994) 67
  [hep-ph/9404201].
  %%CITATION = NUPHA,B425,67;%%

\bibitem{bir}
  J.P.~Blaizot, E.~Iancu and A.~Rebhan,
  %``On the apparent convergence of perturbative QCD at high temperature,''
  Phys.\ Rev.\ D {68} (2003) 025011
  [hep-ph/0303045].
  %%CITATION = HEP-PH 0303045;%%

\bibitem{adjoint}
  K.~Kajantie, M.~Laine, K.~Rummukainen and M.~Shaposhnikov,
  % {\it 3d SU(N) + adjoint Higgs theory and finite-temperature QCD,}
  Nucl.\ Phys.\ B {503} (1997) 357
  [hep-ph/9704416].
  %%CITATION = HEP-PH 9704416;%%

%\cite{Drummond:1997cw}
\bibitem{Drummond:1997cw}
  I.T.~Drummond, R.R.~Horgan, P.V.~Landshoff and A.~Rebhan,
  %``Foam diagram summation at finite temperature,''
  Nucl.\ Phys.\ B { 524} (1998) 579
  [hep-ph/9708426].
  %%CITATION = HEP-PH 9708426;%%

\bibitem{laporta}
  S.~Laporta,
  %``High-precision calculation of multi-loop Feynman integrals by  difference
  %equations,''
  Int.\ J.\ Mod.\ Phys.\  A {15} (2000) 5087
  [hep-ph/0102033];
  %%CITATION = IMPAE,A15,5087;%%
%
  %``High-precision epsilon expansions of massive four-loop vacuum bubbles,''
  Phys.\ Lett.\  B {549} (2002) 115
  [hep-ph/0210336].
  %%CITATION = PHLTA,B549,115;%%

\bibitem{sv}
  Y.~Schr\"oder and A.~Vuorinen,
 %``High-precision evaluation of four-loop vacuum bubbles in three
  %dimensions,''
  hep-ph/0311323;
  %%CITATION = HEP-PH/0311323;%%
%
  %``High-precision epsilon expansions of single-mass-scale four-loop vacuum
  %bubbles,''
  JHEP {06} (2005) 051
  [hep-ph/0503209];
  %%CITATION = JHEPA,0506,051;%%
%
  K.G.~Chetyrkin, M.~Faisst, C.~Sturm and M.~Tentyukov,
  %``e-finite basis of master integrals for the integration-by-parts method,''
  Nucl.\ Phys.\  B {742} (2006) 208
  [hep-ph/0601165].
  %%CITATION = NUPHA,B742,208;%%

%
%


%\cite{Bugrii:1995vn}
\bibitem{Bugrii:1995vn}
  A.I.~Bugrii and V.N.~Shadura,
  %``Three-loop contributions to the free energy of $\lambda\varphi~4$ QFT,''
  hep-th/9510232.
  %%CITATION = HEP-TH 9510232;%%

%\cite{Karsch:1997gj}
\bibitem{Karsch:1997gj}
  F.~Karsch, A.~Patk\'os and P.~Petreczky,
  %``Screened perturbation theory,''
  Phys.\ Lett.\ B { 401} (1997) 69
  [hep-ph/9702376].
  %%CITATION = HEP-PH 9702376;%%

%\cite{Reinbach:1997vi}
\bibitem{Reinbach:1997vi}
  J.~Reinbach and H.~Schulz,
  %``Resummation of phi**4 free energy up to an arbitrary order,''
  Phys.\ Lett.\ B { 404} (1997) 291
  [hep-ph/9703414].
  %%CITATION = HEP-PH 9703414;%%

%\cite{Yamada:1997jw}
\bibitem{Yamada:1997jw}
  H.~Yamada,
  %``Modified Laplace transformation method at finite temperature:  Application
  %to infrared problems of N component phi**4 theory,''
  Int.\ J.\ Mod.\ Phys.\ A { 13} (1998) 4133
  [hep-th/9704001].
  %%CITATION = HEP-TH 9704001;%%

%\cite{Kastening:1997rg}
\bibitem{Kastening:1997rg}
  B.M.~Kastening,
  %``Perturbative finite-temperature results and Pad\'e approximants,''
  Phys.\ Rev.\ D {56} (1997) 8107
  [hep-ph/9708219].
  %%CITATION = HEP-PH 9708219;%%

%\cite{Hatsuda:1997wf}
\bibitem{Hatsuda:1997wf}
  T.~Hatsuda,
  %``Pad\'e improvement of the free energy in high temperature QCD,''
  Phys.\ Rev.\  D {56} (1997) 8111
  [hep-ph/9708257].
  %%CITATION = PHRVA,D56,8111;%%

%\cite{Andersen:1997zx}
\bibitem{Andersen:1997zx}
  J.O.~Andersen,
  %``The screening mass squared in hot scalar theory to order g**5 using
  %effective field theory,''
  Phys.\ Rev.\ D { 57} (1998) 5004
  [hep-ph/9708276].
  %%CITATION = HEP-PH 9708276;%%

%\cite{Leupold:1998yx}
\bibitem{Leupold:1998yx}
  S.~Leupold,
  %``Resummation of soft modes in the free energy of phi**4 theory,''
  hep-ph/9808424.
  %%CITATION = HEP-PH 9808424;%%

%\cite{Andersen:2000zn}
\bibitem{Andersen:2000zn}
  J.O.~Andersen, E.~Braaten and M.~Strickland,
  %``The massive thermal basketball diagram,''
  Phys.\ Rev.\ D { 62} (2000) 045004
  [hep-ph/0002048];
  %%CITATION = HEP-PH 0002048;%%
%\cite{Andersen:2000yj}
%\bibitem{Andersen:2000yj}
  %J.O.~Andersen, E.~Braaten and M.~Strickland,
  %``Screened perturbation theory to three loops,''
  Phys.\ Rev.\ D { 63} (2001) 105008
  [hep-ph/0007159];
  %%CITATION = HEP-PH 0007159;%%
%\cite{Andersen:2001ez}
%\bibitem{Andersen:2001ez}
  J.O.~Andersen and M.~Strickland,
  %``Mass expansions of screened perturbation theory,''
  Phys.\ Rev.\ D {64} (2001) 105012
  [hep-ph/0105214].
  %%CITATION = HEP-PH 0105214;%%

%\cite{Blaizot:2000fc}
\bibitem{Blaizot:2000fc}
  J.P.~Blaizot, E.~Iancu and A.~Rebhan,
  %``Approximately self-consistent resummations for the thermodynamics of  the
  %quark-gluon plasma. I: Entropy and density,''
  Phys.\ Rev.\ D {63} (2001) 065003
  [hep-ph/0005003];
  %%CITATION = HEP-PH 0005003;%%
%\cite{Blaizot:2003tw}
%\bibitem{Blaizot:2003tw}
  %J.P.~Blaizot, E.~Iancu and A.~Rebhan,
  %``Thermodynamics of the high-temperature quark gluon plasma,''
  hep-ph/0303185.
  %%CITATION = HEP-PH 0303185;%%

%\cite{Peshier:2000hx}
\bibitem{Peshier:2000hx}
  A.~Peshier,
  %``HTL resummation of the thermodynamic potential,''
  Phys.\ Rev.\ D {63} (2001) 105004
  [hep-ph/0011250].
  %%CITATION = HEP-PH 0011250;%%

%\cite{Braaten:2001vr}
\bibitem{Braaten:2001vr}
  E.~Braaten and E.~Petitgirard,
  %``Solution to the 3-loop Phi-derivable approximation for massless scalar
  %thermodynamics,''
  Phys.\ Rev.\ D {65} (2002) 085039
  [hep-ph/0107118].
  %%CITATION = HEP-PH 0107118;%%

%\cite{Cvetic:2002ju}
\bibitem{Cvetic:2002ju}
  G.~Cveti\v{c} and R.~K\"ogerler,
  %``Resummations of free energy at high temperature,''
  Phys.\ Rev.\ D { 66} (2002) 105009
  [hep-ph/0207291].
  %%CITATION = HEP-PH 0207291;%%

%\cite{Andersen:2004fp}
\bibitem{Andersen:2004fp}
  J.O.~Andersen and M.~Strickland,
  %``Resummation in hot field theories,''
  Annals Phys.\  { 317} (2005) 281
  [hep-ph/0404164].
  %%CITATION = HEP-PH 0404164;%%

%\cite{Jakovac:2004ua}
\bibitem{Jakovac:2004ua}
  A.~Jakov{\'a}c and Z.~Sz{\'e}p,
  %``Renormalization and resummation in finite temperature field theories,''
  Phys.\ Rev.\ D { 71} (2005) 105001
  [hep-ph/0405226].
  %%CITATION = HEP-PH 0405226;%%

%\cite{Villavicencio:2006vh}
\bibitem{Villavicencio:2006vh}
  C.~Villavicencio and C.A.A.~de Carvalho,
  %``Diagrammatics of the dimensionally reduced action in phi**4 theory,''
  hep-ph/0607250.
  %%CITATION = HEP-PH 0607250;%%

%\cite{Litim:2006ag}
\bibitem{Litim:2006ag}
  D.F.~Litim and J.M.~Pawlowski,
  %``Non-perturbative thermal flows and resummations,''
  JHEP {11} (2006) 026
  [hep-th/0609122].
  %%CITATION = HEP-TH 0609122;%%

%\cite{Blaizot:2006rj}
\bibitem{Blaizot:2006rj}
  J.P.~Blaizot, A.~Ipp, R.~M\'endez-Galain and N.~Wschebor,
  %``Perturbation theory and non-perturbative renormalization flow in scalar
  %field theory at finite temperature,''
  Nucl.\ Phys.\  A {784} (2007) 376
  [hep-ph/0610004].
  %%CITATION = HEP-PH 0610004;%%


  


\end{thebibliography}
\end{document}